\documentclass[acmlarge,dvipsnames,screen]{acmart}

\usepackage{caption}
\usepackage{subcaption}
\usepackage{enumitem}
\usepackage{algpseudocode} 
\usepackage{array}
\usepackage{makecell}
\usepackage{textgreek}
\usepackage{bbding}
\usepackage{changepage}
\usepackage{bmpsize}
\usepackage{pifont}
\usepackage{wasysym}
\usepackage{newtxmath}

\newcolumntype{P}[1]{>{\centering\arraybackslash}p{#1}}
\newcolumntype{M}[1]{>{\centering\arraybackslash}m{#1}}

\newcommand{\xref}[1]{\S\ref{#1}}
\newcommand{\gcheck}[1]{\color{ForestGreen} \ding{52}}
\newcommand{\rcross}[1]{\color{red} \ding{55}}

\AtBeginDocument{%
  \providecommand\BibTeX{{%
    \normalfont B\kern-0.5em{\scshape i\kern-0.25em b}\kern-0.8em\TeX}}}

\newcommand{\red}[1]{\textcolor{black}{#1}}

\setcopyright{rightsretained}
\acmJournal{IMWUT}
\acmYear{2022} \acmVolume{6} \acmNumber{1} \acmArticle{3} \acmMonth{3} \acmPrice{}\acmDOI{10.1145/3517256}

\begin{document}



% \settopmatter{printacmref=false,printfolios=false}
% \setcopyright{acmcopyright}

% \copyrightyear{2021} 
% \acmYear{2021} 
% \acmConference[Proc. ACM Interact. Mob. Wearable Ubiquitous Technol.]{}{Vol. 5, No. 4}{Article 2340}

% \title{Repurposing Smartphone LiDAR for Testing a Drop of Liquid

\title{Testing a Drop of Liquid Using  Smartphone LiDAR}

\author{Justin Chan}
\affiliation{
\institution{Paul G. Allen School of Computer Science and Engineering, University of Washington}
\country{USA}
}
\email{jucha@cs.washington.edu}

\author{Ananditha Raghunath}
\affiliation{
\institution{Paul G. Allen School of Computer Science and Engineering, University of Washington}
  \country{USA}
}
\email{araghu@cs.washington.edu}

\author{Kelly E. Michaelsen }
\affiliation{
\institution{Department of Anesthesiology \& Pain Medicine, University of Washington}
\country{USA}
}
\email{kellyem@uw.edu}

\author{Shyamnath Gollakota}
\affiliation{
\institution{Paul G. Allen School of Computer Science and Engineering, University of Washington}
\country{USA}
}
\email{gshyam@cs.washington.edu}

\renewcommand{\shortauthors}{Chan et al.}

\begin{abstract}
    We present the first system to determine fluid properties using the LiDAR sensors present on modern smartphones. Traditional methods of measuring properties like viscosity require expensive laboratory equipment or a relatively large amount of fluid. In contrast, our smartphone-based method is accessible, contactless and works with just a single drop of liquid. Our design works by targeting a coherent LiDAR beam from the phone onto the liquid. Using the phone's camera, we capture the characteristic laser speckle pattern that is formed by the interference of light reflecting from light-scattering particles. By correlating the fluctuations in speckle intensity over time, we can characterize the Brownian motion within the liquid which is correlated with its viscosity. The speckle pattern can be captured on a range of phone cameras and does not require external magnifiers. Our results show that we can distinguish between different fat contents as well as identify adulterated milk. Further, algorithms can classify between ten different liquids using the smartphone LiDAR speckle patterns.  Finally, we conducted a clinical study with  whole blood samples across 30 patients  showing that our approach can  distinguish between coagulated and uncoagulated blood using a single drop of blood. 
\end{abstract}


\newcommand{\squishlist}{\begin{itemize}[itemsep=1pt,parsep=2pt,topsep=3pt,partopsep=0pt,leftmargin=0em, itemindent=1em,labelwidth=1em,labelsep=0.5em]}
\newcommand{\squishend}{\end{itemize}}
\newcommand{\squishenum}{\begin{enumerate}[itemsep=1pt,parsep=2pt,topsep=3pt,partopsep=0pt,leftmargin=0em,listparindent=1.5em,labelwidth=1em,labelsep=0.5em]}
\newcommand{\squishsubenum}{\begin{enumerate}[itemsep=1pt,parsep=2pt,topsep=0pt,partopsep=0pt,leftmargin=0em,listparindent=1.5em,labelwidth=1em,labelsep=0.5em]}
\newcommand{\squishenumend}{\end{enumerate}}

\settopmatter{printfolios=true}
\maketitle

\title{Testing a Drop of Liquid Using Smartphone LiDAR}

\section{Introduction}

\begin{figure}[t]
  \centering
    \includegraphics[width=.4\textwidth]{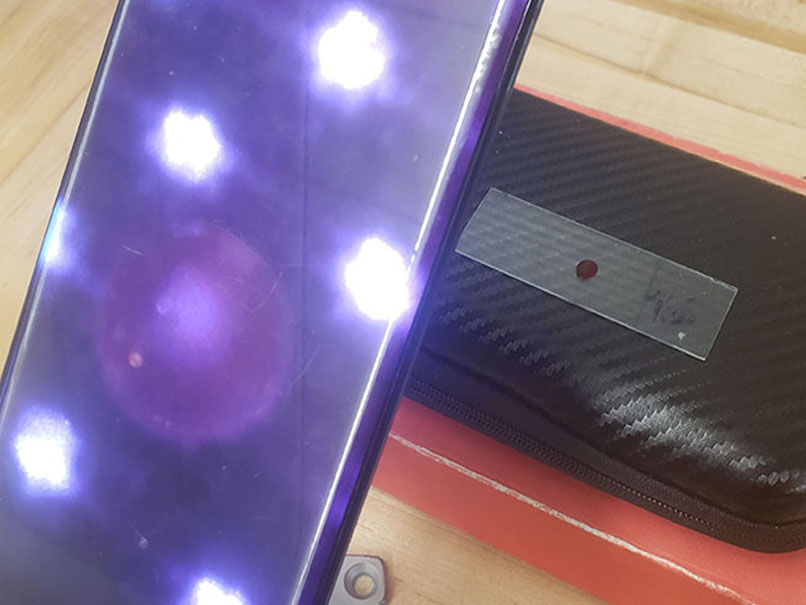}
    \includegraphics[width=.4\textwidth]{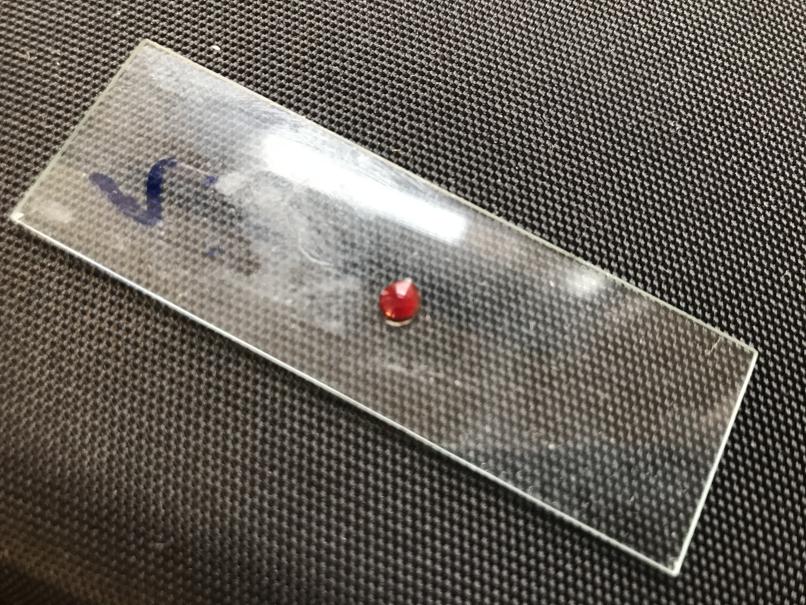}
\caption{Generating laser speckle using smartphone LiDAR. (Left) Smartphone LiDAR transmissions  as captured by a modified  smartphone camera, (Right) close-up of 10 \textmu l of whole blood being tested on a glass slide.}
\label{fig:fig1}
\end{figure}

Smartphones are ubiquitous devices that come with a powerful set of sensors ranging from microphones, speakers, GPS, accelerometers, cameras and touch sensors. 
While these sensors are powerful as standalone components, they have also been repurposed and combined in innovative ways to enable various applications spanning  human computer interaction~\cite{hci1,hci2,hci3}, mobile health~\cite{health1,health2,capcam} and wireless  communication~\cite{comm1}. 

The latest sensor addition is the light detection and ranging (LiDAR) scanner that has been shipping in new iPhone models starting late 2020. Scanning LiDAR systems transmit a train of laser pulses directed at different parts of a scene over a short period of time. These near-infrared laser pulses bounce off objects in the scene and return to the sensor, which computes the time of flight to estimate the distance and depth of various objects in the scene. While augmented and virtual reality are the primary use cases for these smartphone LiDAR sensors, they also provide a  coherent and targeted laser source that can be repurposed for sensing applications.

In this paper, we explore the idea of repurposing the smartphone LiDAR to sense physical properties (e.g., viscosity) from a drop of liquid. Liquid sensing using smartphones can enable an accessible and contactless tool with applications in both biomedical sensing and food rheology. In biomedical sensing, determining the coagulation state from a single drop of blood {(a  single drop of liquid has a volume of around 10~\textmu l~\cite{papercentrifuge}. It is also about the quantity of blood that a patient can collect using a finger lancet device)} can help millions of people with increased risk of morbidity and mortality from blood clotting disorders~\cite{authors2006acc}. Frequent blood coagulation testing is critical for these patients but testing is currently performed in laboratories or with expensive point-of-care devices that limit test frequency and affordability~\cite{cite1,cite4,cite5,cite6}. In food rheology, the fat concentration and physico-chemical properties associated with fat globules and protein affect viscosity of foods (e.g., milk)~\cite{obesity}. Thermal and mechanical operations can modify the rheological characteristics of these compounds~\cite{dairylaser}. A smartphone-based approach to determining food rheology can be an important and accessible quality control tool in food manufacturing and processing.

Prior work on liquid sensing uses radio signals to identify the properties of liquids (e.g., permittivity). These systems either use phase and amplitude changes in the radio signals transmitted from custom  ultra-wide band radios and radar hardware~\cite{liquid,radarcat} or radio coupling with RFID tags attached to the container~\cite{rfiq,tagscan}. More recent work uses the vibration motor on the smartphone to create capillary waves in the liquid to compute its surface tension~\cite{capcam_0,capcam}.  However these systems require that the depth of the liquid in a standard paper cup is more than 25 mm~\cite{capcam_0}. This is orders of magnitude more liquid than a single drop (10 \textmu l). Collecting such liquid volumes can be challenging in our target applications like home-testing of blood coagulation.

Our work instead leverages the {\it laser speckle phenomena}~\cite{abou2016evaluation}. When highly coherent laser light is illuminated on a diffuse surface, it produces a constructive and destructive interference pattern known as a speckle  (see Fig.~\ref{fig:fig2}). When the variation in the surface height is larger than the laser wavelength, light from different points on the surface within the camera's pixel resolution traverses different paths and hence superpose with each other to create constructive and destructive interference patterns. In the context of liquids, when minute particles that are larger in size than the laser wavelength ({\char`\~ 800~nm ~\cite{forbes}}) move within the liquid, the speckle pattern changes with time. Since the Brownian motion of particles (e.g., platelets) in a liquid is affected by its viscosity, we can get a proxy measure of the liquid viscosity by analyzing the time-varying speckle.

To achieve this, we repurpose the LiDAR hardware in smartphones as a source of highly coherent, single-frequency laser light. The smartphone LiDAR sends out a mesh of near infrared laser pulses in a grid formation of 24 by 24 points. We place the drop of liquid (e.g., blood) on a glass slide and position it to overlap with one of the laser points as shown in Fig.~\ref{fig:fig1}. We then use a smartphone camera to analyze the time-varying speckle phenomena. Since the laser transmissions are in the near-infrared range, we manually remove the filter on the camera to capture the near-infrared spectrum. {Note that the iPhone has an onboard near infrared camera that can analyze the LiDAR transmissions. However,  the raw images from the near infrared camera are not yet made available in software to third-party developers. To be practical, our techniques need to   be implemented in software using the onboard near infrared camera by smartphone manufacturers so that we do not need to physically removing the filter on the RGB camera of the phone risking damage or use a cheaper smartphone as the receiver.} By analyzing the resultant time-varying laser speckle observed by the camera, we infer information about the liquid.

Repurposing the LiDAR hardware on smartphones for laser speckle reflectometry requires addressing multiple technical challenges. In contrast to a continuous wave transmission from conventional lasers, the LiDAR transmitters on the iPhone  use pulse width modulation to duty-cycle the transmissions on and off at a high frequency. This makes it challenging for a receiving camera to be able to capture a stable image of the speckle reflections. Instead, the receiving camera will observe distortions of the LiDAR transmissions as a result of the rolling shutter effect. This manifests as three different types of distortions. The first distortion is a high frequency on-off flickering of the LiDAR transmissions. The second distortion is a lower frequency pattern of black bars that move across the screen and periodically obscure the transmissions. The final distortion is the occasional skewing and wobbling on the transmissions. This occurs when CMOS image sensors, the predominant image sensor on smartphones, capture fast changes in lighting, scanning across an image vertically or horizontally, as opposed to capturing the entire frame simultaneously as is the case with CCD image sensors.

\begin{figure}[t]
\vskip 0.01in
  \centering
    \includegraphics[width=.32\textwidth]{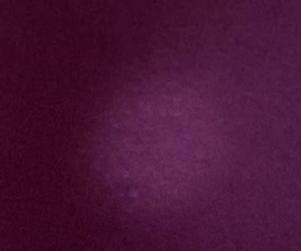}
    \includegraphics[width=.32\textwidth]{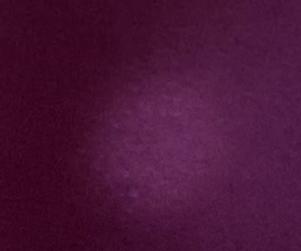}
    \includegraphics[width=.32\textwidth]{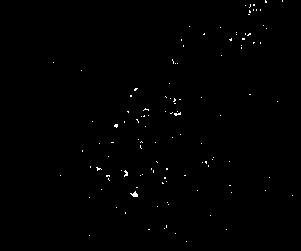}
% \vspace{-0.1in}
\caption{Closeup of laser speckle captured at frame $i$ (left) and frame $i+9$ (middle) for a 10 \textmu l sample of whole blood using smartphone LiDAR. The rightmost image shows the difference in between the two images. This time-varying pattern is caused by the Brownian motion of particles (e.g., platelets) in the liquid.}
% \vskip -0.1in
\label{fig:fig2}
\end{figure}

We design algorithms to address the above challenges and implement our design on off-the-shelf smartphones. We run benchmark experiments to understand the effects of various design parameters like camera shutter speed and zoom settings, distance from the sample, liquid volume, background light and the surface material the sample is place on. Our experiments show that our design does not require precise liquid measurements and can operate with volumes between 10--50~\textmu l. Further, our algorithms can operate at shutter speeds of 1/30 and 1/60s, with a frame rate of 30 frames per second and account for the rolling shutter effect. 

Building on these benchmarks, we evaluate our design in three  applications. 
\squishlist
\item {\bf Blood coagulation detection.}
We conducted a clinical study using 30 anonymized whole blood samples from the  medical center in our institution with samples from the anticoagulation clinic. Our results show that laser speckle generated by the smartphone LiDAR can accurately distinguish between coagulated and uncoagulated blood. Further, this can be achieved using a single drop (10~\textmu l) of whole blood.
\item {\bf Milk sensing.} The second application is to use the smartphone LiDAR to track fat contents in milk and to  identify adulterated milk. Our results show that we can differentiate between 0\%, 1\%, 2\% fat milk, whole milk and cream.  Further, we can differentiate whole milk from milk adulterated with substances like detergent, cornstarch and xanthan gum.

\item{\bf Laser-speckle based liquid classification.} We train a support vector machine to classify between the laser speckle patterns (i.e., the black and white images in Fig. 2) for 10 different liquids ranging from sparkling water, coffee, vinegar and dish soap.  Our classification could classify between the 10 classes with an average accuracy of 91.5\% on test data corresponding to  independent measurements from the training data.
\squishend

\vskip 0.05in\noindent{\bf Limitations.} To ensure consistency across experiments, we tested all liquid samples at room temperature in an area with the same relative humidity level. These conditions are necessary as the viscosity of a liquid can change in response to environmental temperature. The liquid samples also remained stationary on the glass slide as movements to the setup would interfere with the speckle pattern and preclude the measurement from meaningful analysis. We envision that our smartphone LiDAR technique would be useful as a screening tool to detect unknown adulterants in liquids (e.g., milk) by identifying deviations from known viscosity values. With this technique, identifying the chemical composition of an adulterant is challenging, and further testing might be required to measure chemical composition of unknown adulterants. Finally, we leverage a machine learning-based approach for liquid classification instead of obtaining a closed-form equation to calculate viscosity across different classes of liquids. This is because the particles that create the laser speckle pattern differ depending on the class of liquid being tested. For example, while the speckle pattern from blood is caused by the Brownian motion of red blood cells and platelets, the pattern in milk is caused by fat globules and large proteins. As such, deriving a closed-form equation is challenging, and using machine learning  is a more compelling approach.

\vskip 0.05in\noindent{\bf Contributions.} While prior work has used the phenomena of laser speckle using custom hardware~\cite{lasercapp2,lasercapp3,lasercapp4,lasercapp5}, this paper makes five key contributions: 1) We introduce the idea of repurposing the LiDAR hardware on smartphones to enable laser speckle reflectometry and demonstrate its use for liquid sensing. 2) We present algorithms to extract the speckle information from the duty-cycled pulse width modulated laser transmissions from the smartphone and address issues such as the rolling shutter effect. 3) We perform a clinical study to show the feasibility of differentiating between coagulated and uncoagulated blood using the laser speckle patterns from the smartphone LiDAR. 4) We  show that our approach can be used to perform milk sensing, in particular, for tracking fat content and identifying adulterated milk. 5) Finally, we  show that our smartphone-based approach can be used to classify between liquids by training a support vector machine across ten different liquids spanning a range of viscosities.

\section{Laser Speckle Background}

Laser speckle reflectometry uses reflections from coherent laser transmissions. Unlike  light transmitted from LEDs, coherent light refers to a stream of photons that are of a single frequency and have the same phase difference. Laser is a form of light that is both spatially and temporally coherent, in that it can be focused into a narrow beam and emit a narrow wavelength spectrum. When a laser light is initially transmitted, it is coherent, but when it impinges upon a medium, it scatters in different directions due to minute variations in surface roughness. The scattered waves interfere constructively and destructively to create a distinct pattern of bright and dark points respectively known as laser speckle. This phenomena occurs when the roughness of the surface is comparable to or larger than the wavelength of the impinging laser (in the case of near-infrared transmissions from smartphone LiDAR, this is around {800~nm~\cite{forbes})}. 

{\it Static speckle.} When a laser is directed at a static rough surface, the resulting speckle pattern distinctly corresponds to the texture of the surface~\cite{briers2013laser}. Such a pattern is known as an objective speckle as it depends only on the wavelength of the laser and the object under inspection, not on the imaging system. These static speckle patterns do change with time for solid objects and can be used to identify different textures~\cite{laserplastics}. 

{\it Dynamic speckle.} When the laser is directed towards a liquid, it will reflect off light-scattering particles in the liquid and create dynamic speckle patterns that change due to the motion of particles. For example, when a laser is pointed at a sample of milk, particles in the form of globs of fat or protein scatter the light and produce a speckle pattern. The speckle pattern is very sensitive to changes in displacement of these particles. So when these particles move and fluctuate in accordance to Brownian motion, the appearance of the speckle pattern will change correspondingly. %This allows for time-series laser speckle analysis.  

Prior work has shown that the rate of change of Brownian motion in these particles is correlated with its viscosity~\cite{abou2016evaluation}. In other words, particles in liquids that are viscous move at a comparatively slower pace compared to particles in liquids that are not as viscous. For example, when a laser is directed at a clotting blood sample, the interactions between fibrin and platelets affect the Brownian motion of light-scattering particles. As the clot starts to form the amount of particle Brownian motion  decreases, and the speckle pattern changes more slowly. By calculating the rate of change of speckle pattern, one can estimate a viscosity coefficient of a liquid, and track viscosity changes over time.

\section{Our design}
We first describe the challenges with using the LiDAR hardware on smartphones for laser speckle reflectometry. We then describe the processing pipeline to extract the speckle pattern using the smartphone LiDAR. Finally, we describe the techniques we use to address practical issues such as the flickering effect.

\begin{figure}[t]
\centering
\includegraphics[width=.48\textwidth]{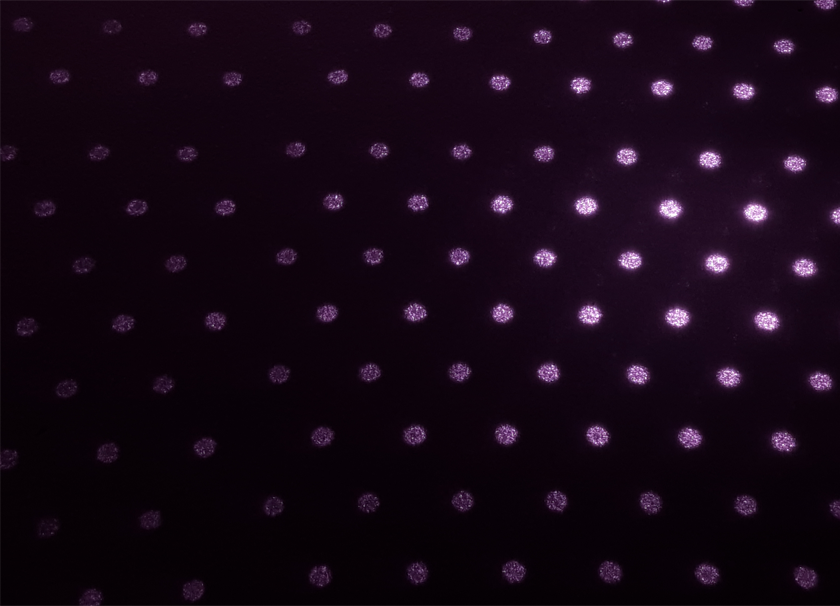}
\includegraphics[width=.48\textwidth]{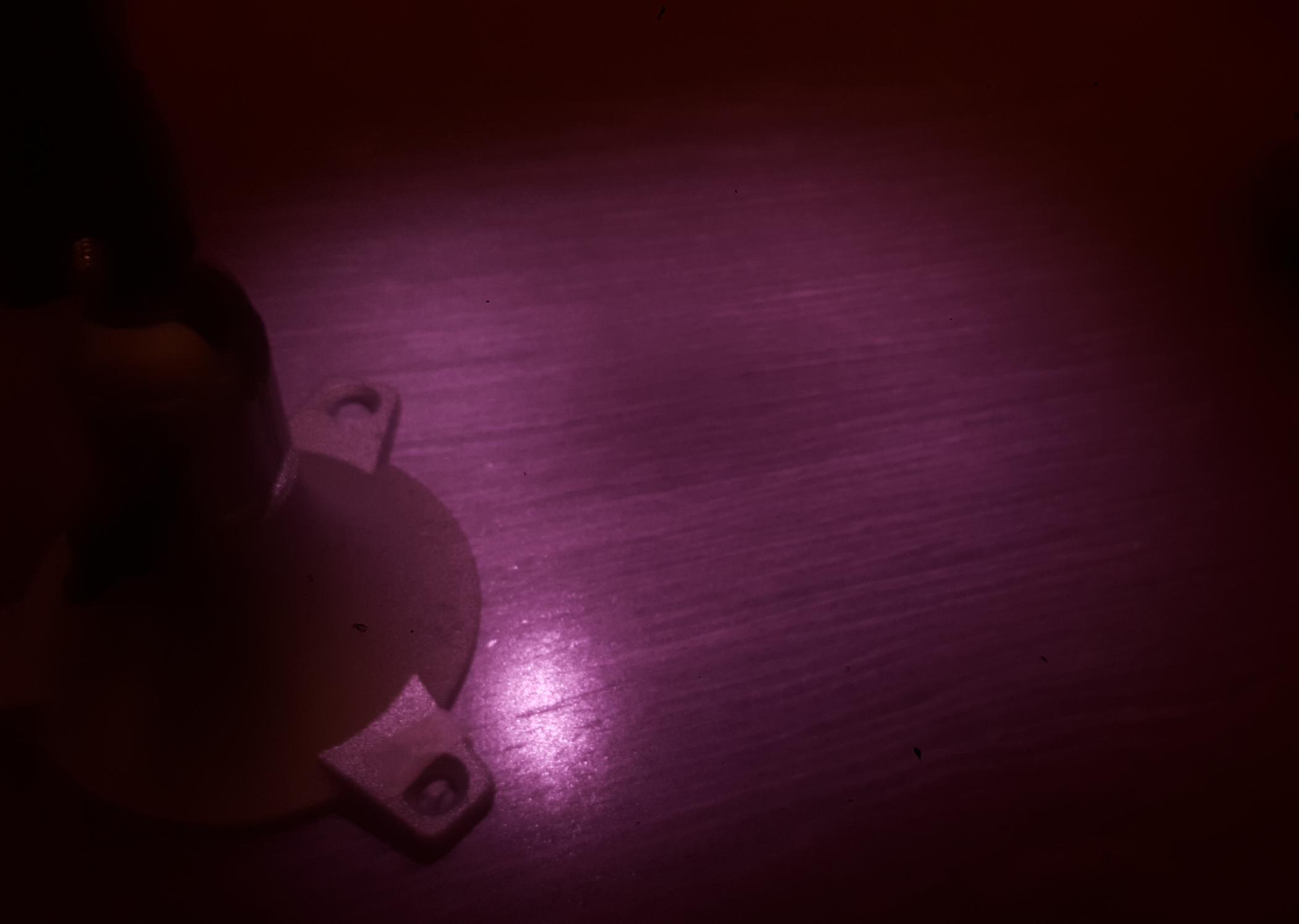}
% \vspace{-0.1in}
\caption{(Left) Focused LiDAR dot projections from the iPhone 12 and (Right) diffuse laser projection from the Samsung Galaxy Note 10+. The dot projections have a higher power and are more focused allowing for easier speckle pattern extraction.}
\label{fig:phone_speckle}
\end{figure}

\subsection{Smartphone LiDAR Challenges}
\label{sec:challenges}

In this section we describe the landscape of LiDAR sensors on smartphones and the challenges of using these transmissions for sensing applications that rely on laser speckle.

{\it iPhone devices.} By market share, the most prominent phones to be equipped with LiDAR sensors are the iPhone 12 Pro and iPhone 12 Pro Max. They have a LiDAR projector and time-of-flight camera on the rear of the phone. The projector sends out a mesh of near infrared laser pulses in a grid formation of 24 by 24 points as shown in Fig.{~\ref{fig:phone_speckle}a}. This can be used for depth sensing and for augmented reality applications with a working range of up to 5~m. When we place the LiDAR projector at a distance of 5~cm from a surface, the beam width of each projected laser point is about 1~mm in diameter. 

The front of the iPhone 12 Pro/Max is also equipped with a near infrared transmitter and camera known as TrueDepth that is used for Face ID, a facial recognition system for biometric authentication. This transmitter sends out a denser mesh of 30,000 laser points that are designed to create a model of the user's face and is work at an arm's length of 25-50~cm. We note that when the front of the phone is placed close to a surface, the individual laser points are no longer discernible by eye, and instead appear as a single cluster of diffuse near-infrared light {(see Fig.~\ref{fig:phone_speckle}b)}. Additionally, we note that the front-facing Face ID LiDAR is designed for shorter ranges, each individual beam carries less power than the LiDAR transmissions from the rear of the phone. In comparison to the LiDAR project, the lower power along with the diffuse near-infrared light as seen by the camera, makes it challenging to achieve laser speckle imaging using Face ID.

Both the LiDAR transmitters on the iPhone also use pulse width modulation to duty-cycle the transmissions on and off at a high frequency which makes it challenging for a receiving camera to be able to capture a stable image of the transmissions. Instead, the receiving camera will observe distortions of the LiDAR transmissions as a result of the rolling shutter effect. This manifests as three different types of distortions. The first distortion is a high frequency on-off flickering of the LiDAR transmissions. The second distortion is a lower frequency pattern of black bars that move across the screen and periodically obscure the transmissions. The final distortion is the occasional skewing and wobbling on the transmissions. This effect occurs when CMOS image sensors, the predominant image sensor on smartphones, captures fast changes in lighting scanning across an image vertically or horizontally, as opposed to capturing the entire frame simultaneously as is the case with CCD sensors. 

The typical solution to resolve the rolling shutter effect is to increase the camera's shutter speed in order to average out the image distortions over a longer period of time. We find that with a shutter speed of 1/15~s, the receiving camera (with its filters removed) is able to obtain a stable image of LiDAR transmissions from the iPhone's rear transmitter. However, transmissions from the front LiDAR transmitter still create a rolling shutter effect even with high shutter speeds, making it undesirable for use as part of laser speckle based sensing. This effect occurs as the LiDAR transmitter is not synchronized with the receiving camera. Although the iPhone's onboard near-infrared cameras are synchronized to the LiDAR transmissions, the near infrared images are inaccessible to end users. These images are stored within the iPhone's Secure Enclave, a co-processor designed for data protection and privacy. This system is designed such that none of the images are accessible by the phone's main processor, and usage of the camera is not exposed in any of the iPhone's API documents.

Although we are able to capture stable images of LiDAR transmissions from the  filter-modified camera, our frame rate is consequently reduced to 15 frames per second, and each frame is temporally averaged over a longer duration than is typical for laser speckle sensing systems. Typical laser speckle based sensing systems use shutter speeds from 1/750~s or lower in order to more accurately capture the dynamics of laser speckle from light-scattering particles.

{\it Android devices.} There are several other phones that have near-infrared sensors including the Samsung Galaxy Note 10+, the Huawei P30 and the Google Pixel 4 and 4XL. The transmitters on these phones are confirmed to be near infrared laser transmitters (as opposed to LED transmitters) as pointing a near infrared camera at them will cause an objective speckle pattern to appear directly on the camera's image sensor. These phones project a diffuse beam that covers a larger region instead of a single point. Because of this, the amount of laser power at any particular point in space is low, and when these phones are pointed towards a liquid, very little light is scattered from the particles within the liquid and a speckle pattern is not visible. While it is in principle possible to find a focusing lens to direct the beams towards a single point, identifying an appropriate lens and installing it can be an interesting future direction.

We also note that phones such as the Google Pixel 4 and 4XL and the DOOGEE S96 Pro make their near infrared camera accessible to end users, little control is provided over the camera's image settings. Specifically, there is no way to adjust the focus, shutter speed, ISO (brightness) or white balance on the cameras. In addition, there is no way to turn off the camera's automatic ISO adjustment feature. This can result in unwanted saturation of the camera's image if the camera is too close to a surface, which can make it challenging to perform controlled experiments.

\subsection{Smartphone Speckle Processing}
\label{sec:processing}

\begin{figure*}[t]
\includegraphics[width=.8\textwidth]{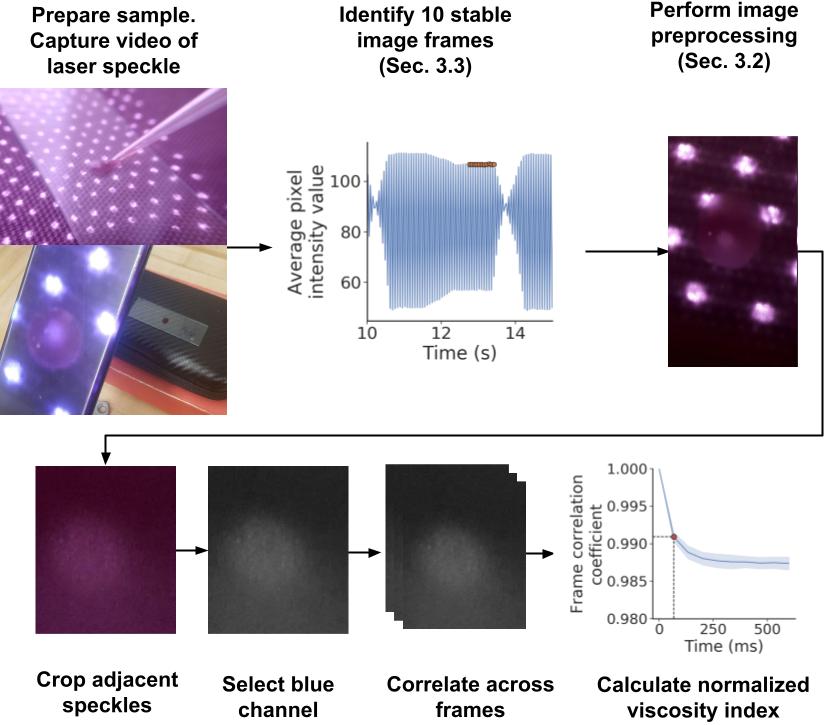}
\caption{Our processing pipeline to extract the normalized viscosity index from the smartphone LiDAR.}
% \vspace{-0.2in}
\label{fig:flow}
\end{figure*}

At a high level, when coherent LiDAR transmissions are pointed towards a drop of liquid, the transmissions are scattered from particles within the liquid and cause a random interference signal in the form of a speckle pattern. The dynamic changes in speckle over time reflect the Brownian motion of particles within the liquid and are correlated with the liquid's viscosity. The amount of speckle movement over time can be computed frame by frame using spatial speckle contrast which is defined as,
$K=\frac{\sigma}{<I>}$. Here $\sigma$ is the standard deviation of speckle intensity in the frame and $<I>$ is the mean image intensity in the frame~\cite{survey}.

In this section, we describe the various processing steps to obtain the viscosity coefficient for a video of smartphone laser speckle being scattered from a drop of fluid.

\vskip 0.05in\noindent{\it(1) Signal pre-processing.} Our goal is to extract a set of video frames of the recorded speckle that are centered on the speckle and that do not capture other sources of image fluctuation in the image. We discard the first and last five seconds of the video as a transient time where the frame could be moving subtly in response to pressing the video record button on the smartphone. Further, we only look at the blue  channel of the RGB image frames. While red is the closest color to the near infrared spectrum, we do not pick it to avoid image saturation  by the co-located LiDAR. 

\vskip 0.05in\noindent{\it(2) Cropping frames.} We are only interested in the dynamic speckle being reflected from the liquid, not other static speckle patterns that may also be within the image. As the videos are recorded so that the speckle is centered in the middle of the frame, we are able to crop out other static speckle patterns. To do this, we apply a color threshold to the blue color channel of the image to only include pixel regions that are greater than 200, these represent other bright static speckles that are in the image. We create a candidate bounding box centered at the center of the image with length and width of 2 pixels. We progressively increase the width and height of the bounding box in steps of 1 pixel, while checking if the bounding box overlaps with any of the static speckles. Once it does, we stop increasing the bounding box size and crop to that region within the image. If the bounding box does not overlap with any static speckles, we terminate the algorithm when the bounding box has grown to 1000 X 1000 pixels.

\vskip 0.05in\noindent{\it(3) Computing correlation curves.} The goal of this step is to calculate the amount of change in the laser speckle pattern between frames. To characterize the rate of change in the laser speckle pattern over time, we perform a two-dimensional correlation analysis between video frames $t$ and $t,t+\tau,t+2\tau,...,t+9\tau$. Specifically, the correlation coefficient between two frames is computed as, 
$$\frac{\sum_{x=0}^{X} \sum_{y=0}^{Y} (I_{xy}(t) - \bar{I}(t)) (I_{xy}(t+\tau) - \bar{I}(t+\tau))}{\sqrt{(\sum_{x,y} (I_{xy}(t) - \bar{I}(t))^2) (\sum_{x,y} (I_{xy}(t+\tau) - \bar{I}(t+\tau))^2) }}$$.

Here $I_{xy}(t)$ is the intensity for the pixel (x,y) at time t and $\bar{I}(t)$ is the intensity at time t averaged across all the pixels. The intuition here is that in the case of rapid Brownian motion of particles, the speckle pattern changes quickly over time, and the video frames would decorrelate quickly. Conversely, if there is little change in particles over time, which would occur in the case of viscous liquids, the video frames would exhibit a higher degree of correlation. 

The result of this correlation analysis is the correlation curve shown in Fig.~\ref{fig:flicker}a. The curves show that the largest dip is between the first and the second frame. As the movement of speckle pattern and the corresponding Brownian motion is random and rapid, the first frame will quickly decorrelate with subsequent frames and reach a plateau within a matter of two to three frames, which indicates that the speckle pattern is now uncorrelated.

\vskip 0.05in\noindent{\it(4) Calculating normalized viscosity coefficient.} From this correlation curve a viscosity coefficient id determined $V$ that is correlated with ground truth viscosity measurements. At a high level, the steepness of the curve represents how quickly the speckle pattern is decorrelating. The more quickly the curve decorrelates, the less viscous the liquid is. We use the second point along the graph to represent the amount of decorrelation between frames, and use that as our normalized viscosity coefficient $V$. We show in~{\xref{sec:milkresults}} that this coefficient correlates with ground truth viscosity measures. We also note that alternatively, an exponential function of the form $e^\frac{-\tau}{\tau_c}$ can be fit to the correlation curves~\cite{nader2017assessing}. The speckle decorrelation time constant $\tau_c$ has been shown  to correlate linearly with liquid viscosity~\cite{abou2016evaluation}.

\begin{figure*}[t]
  \centering
    \includegraphics[width=.32\textwidth]{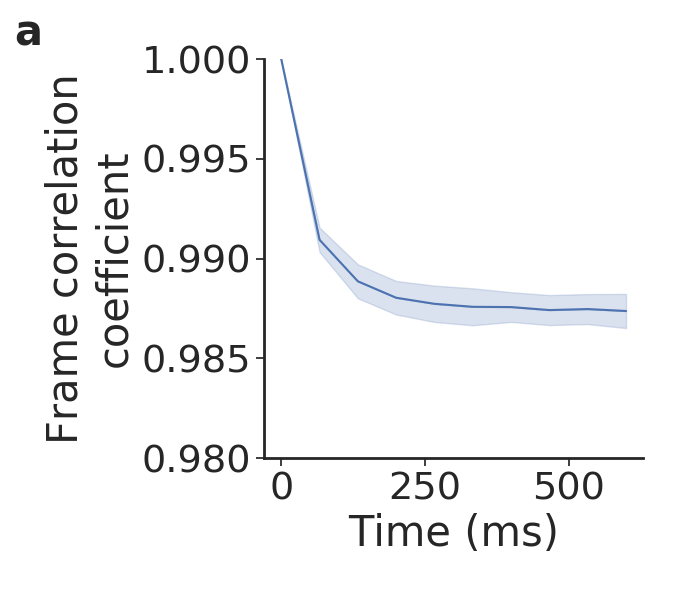}
    \includegraphics[width=.32\textwidth]{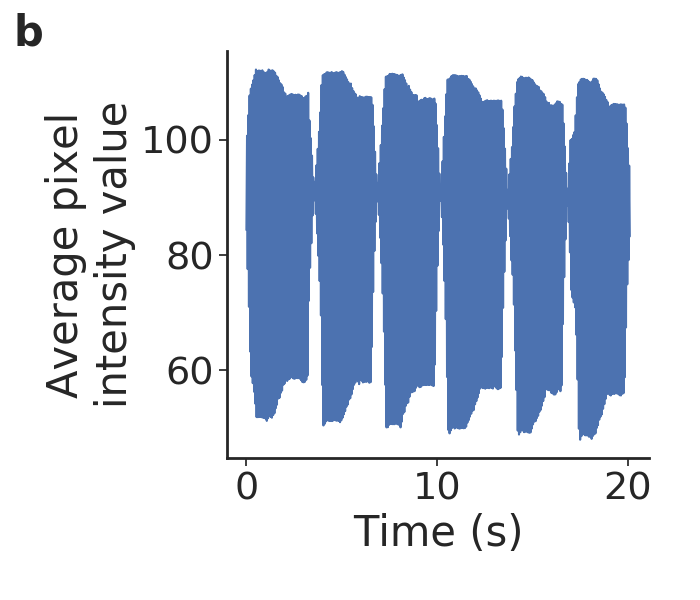}
    \includegraphics[width=.32\textwidth]{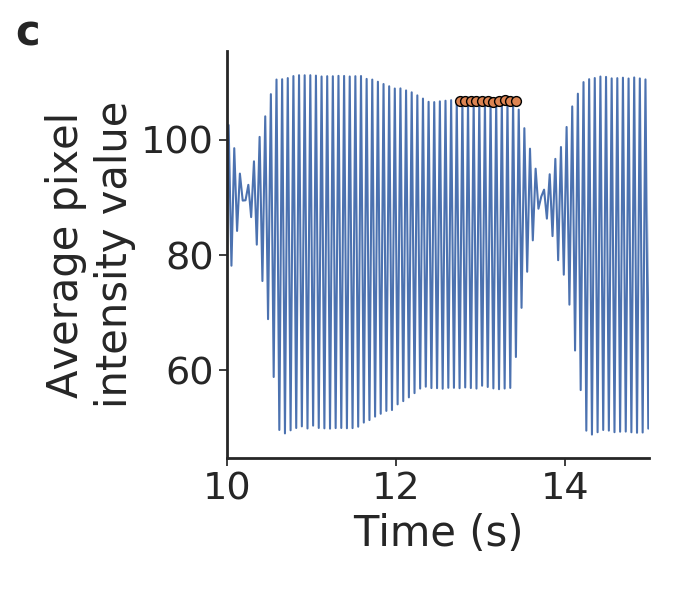}
    \vskip -0.2in
    \caption{(a) Correlation curve for a 20 \textmu l sample of whole milk. (b-c) Average pixel intensity values with a shutter speed of 1/30~s illustrating the artifacts introduced by the rolling shutter effect.}
\label{fig:flicker}
% \vspace{-0.1in}
\end{figure*}

\subsection{Addressing Practical Issues}

As described in~\xref{sec:challenges}, the iPhone LiDAR transmissions are driven using a PWM signal and are duty-cycled on and off at a fast frequency. When the camera captures a video of these transmissions at shutter speeds of 1/30 or 1/60~s, this results in a rolling shutter effect that cause three different types of distortions in the image. In this section we describe the three distortions and our algorithms for addressing these distortion. Our goal is to select ten consecutive laser speckle frames that can be used to generate the correlation curve and estimate the viscosity coefficient. 

\vskip 0.05in\noindent{\it (1) Flickering effect.} Fig.~\ref{fig:flicker}(b,c) plots the average pixel intensity value for each frame across a laser speckle video as a function of time. Fig.~\ref{fig:flicker}(c) shows that there is a high frequency ON-OFF pattern that corresponds to the flickering effect. To address this issue, our algorithm ignores all frames that show up as minimums in the figure as these represent frames where the speckle pattern is not visible. We only select the frames corresponding to the peaks where  the speckle is at least visible for the next step.

\vskip 0.05in\noindent{\it (2) Low frequency bars.} In addition to the high frequency flicking effect, Fig.~\ref{fig:flicker}(b) shows a low  frequency changes over time. Specifically the figure shows a period of five nulls where a low frequency pattern of black bars moves across the screen and periodically obscure the transmissions. To avoid the peaks falling in these nulls from consideration, we normalize the signal between 0 and 1 and remove any peaks that fall below a threshold. In our implementation we use an amplitude threshold of 0.85.
\vskip 0.05in\noindent{\it (3) Skewing and wobbling.} Skewing and wobbling of the image is seen in Fig.~\ref{fig:flicker}(c) as downward or upward slopes in the average pixel intensity values instead of a flat portion of pixel intensity values. This is caused due to the rolling shutter issue. Our observation however is that these rolling shutter issues cause much greater changes than the laser speckle fluctuations. Thus, our goal is to select peaks that fall on a more flat and stable portion of the signal, and avoid selecting peaks that fall along these slopes. To do this, we scan through the entire waveform for windows of ten consecutive peaks, where the range between the peaks is minimized. The window with the minimum range is then passed into the correlation algorithm described in~\xref{sec:processing} and used to calculate a normalized viscosity coefficient.

\begin{figure}[t!]
    \includegraphics[width=.32\textwidth]{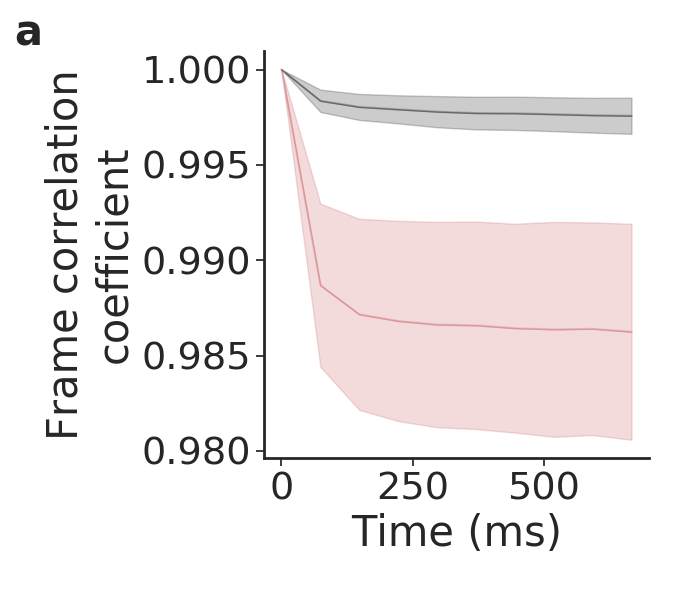}
    \includegraphics[width=.32\textwidth]{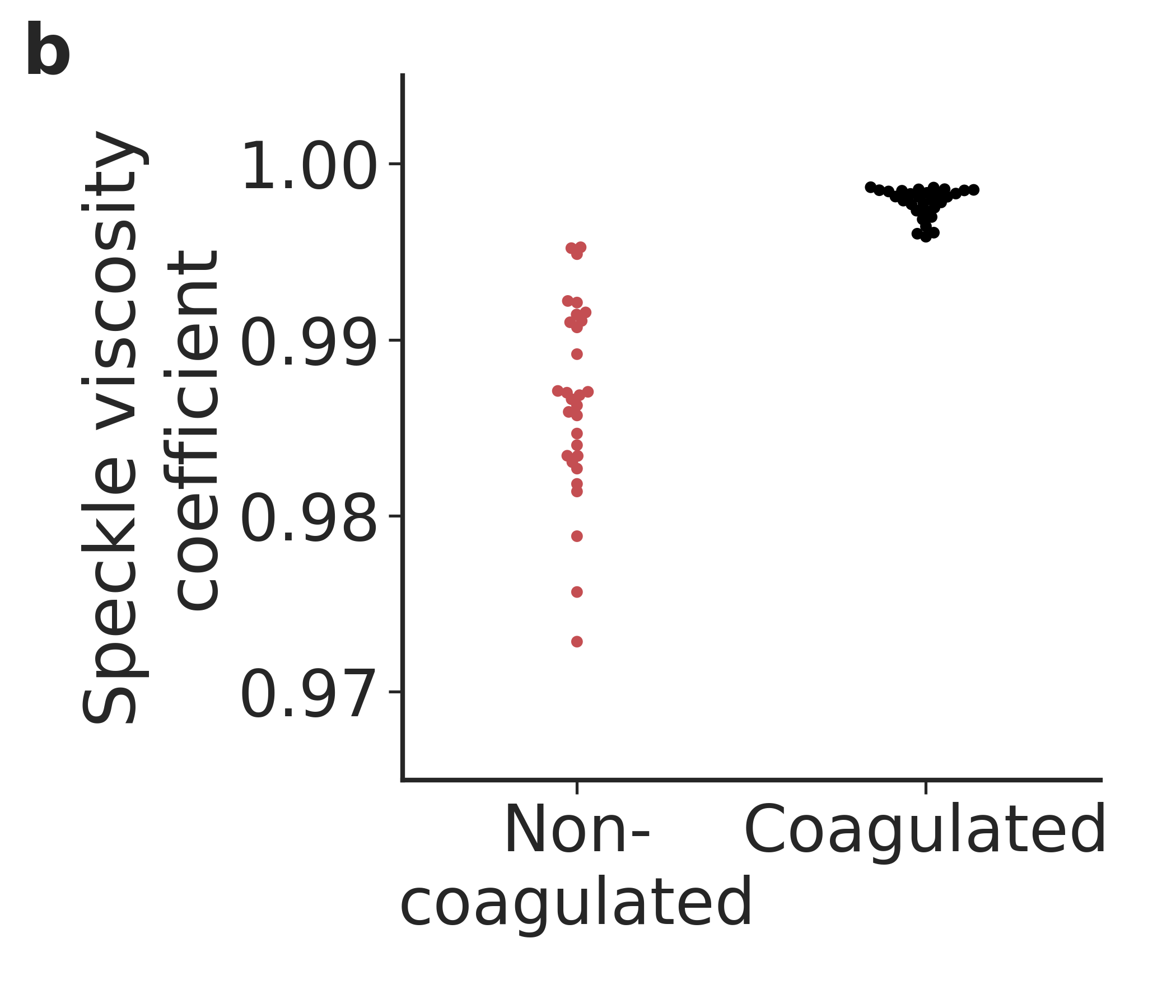}\\
    \includegraphics[width=.64\textwidth]{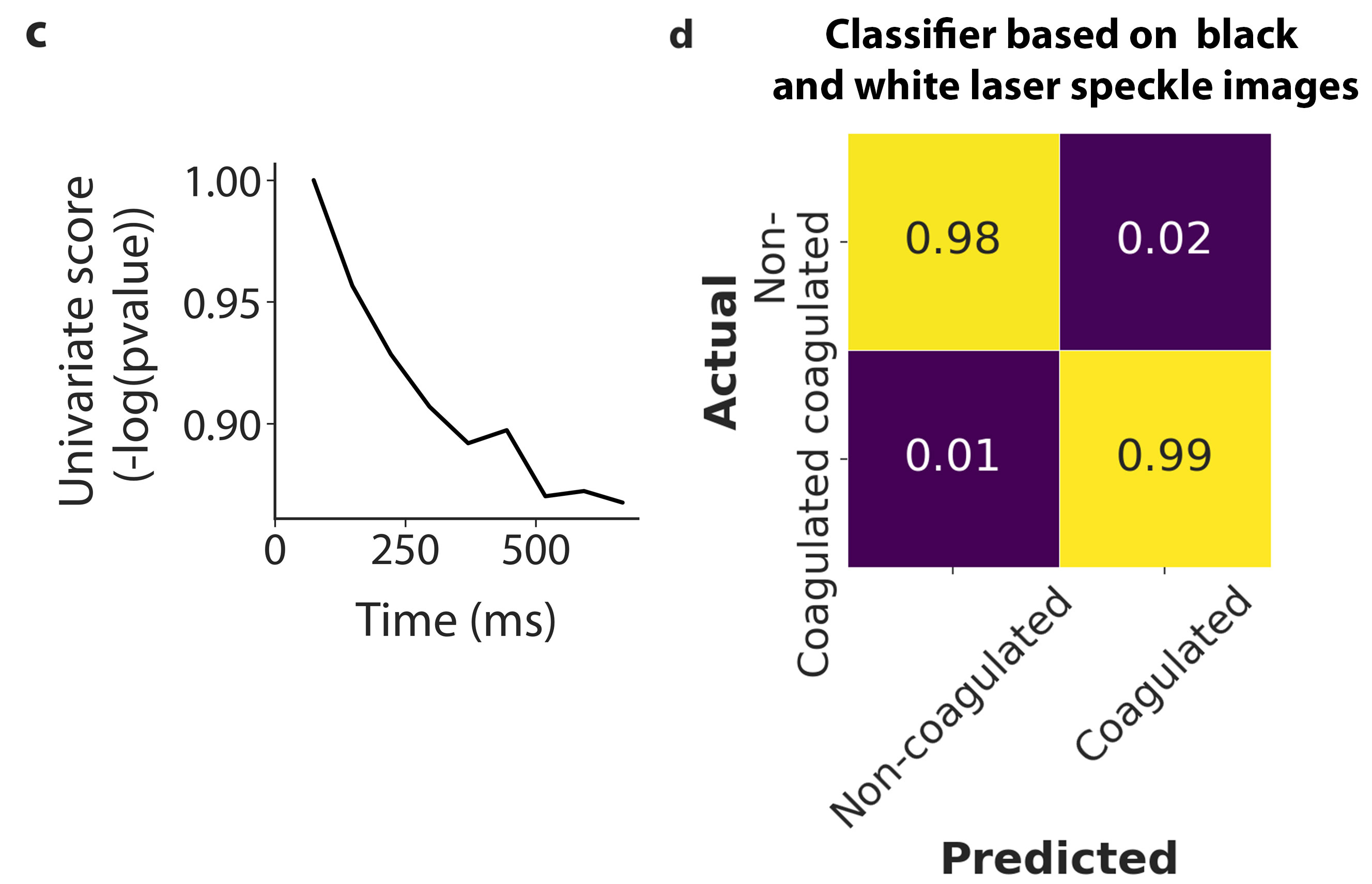}
    \\
\caption{Blood coagulation testing. (a) Correlation curves for the 30 non-coagulated blood samples (red) and the 30 coagulated blood samples (gray). (b) The normalized viscosity coefficients for the 30 coagulated and non-coagulated blood samples. (c) The result of the univariate analysis on the correlation curves from the two classes. (d) Confusion matrix for classifying between non-coagulated and coagulated blood. The 30 different measurements correspond to blood samples from 30 different human participants.}
\label{fig:blood}
% \vspace{-2em}
\end{figure}

\section{Evaluation}

We first evaluate our smartphone LiDAR based speckle design with coagulated and non-coagulated blood. We then evaluate it in our milk sensing application. Finally, we present benchmark experiments to understand the effect of various design parameters like shutter speed and  distance.

\subsection{Blood Coagulation Study}
The first application we explore is differentiating between coagulated and uncoagulated blood. The motivation here is that millions of people suffer from blood clotting disorders such as hemophilia and other coagulopathies that affect the ability for the person's blood to clot. People with these conditions are at an increased health risk as minor incidents of bleeding can result in significant damage. An accurate diagnosis can result in treatment of the disorder and restoration of quality of life. Being able to diagnose people with blood clotting disorders particularly in low resource settings in a timely and efficient manner can have a positive effect on health outcomes~\cite{cite1,cite4,cite5,cite6}. This application also requires that the amount of blood we can operate on is minimized. Ideally, a patient can collect around a drop of blood (around 10~\textmu l) using a finger lancet device.

We designed a clinical study to investigate if our smartphone based laser speckle system is able to differentiate between non-coagulated and coagulated blood. We obtained institutional review board approval to collect de-identified whole blood samples at our institution's medical center. We collected 30 whole blood samples from patients at the medical center, which includes patients at the hospital's anticoagulation clinic. Samples were preserved in a test tube containing citrate to ensure the samples did not coagulate within the tube. Samples were collected and stored in a refrigerator at 4°C to increase their shelf life. Prior to testing each sample, each blood sample which is in a test tube is heated in a water bath to a temperature of 37°C for a period of three minutes to get it to  mimic human body temperature. After this time, 10~µl of the blood sample was pipetted onto a glass slide and placed in view of our smartphone LiDAR transmitter and receiving camera for sensing at a distance of 5~cm. Special care was taken to ensure that no air bubbles had formed in the blood sample upon pipetting. Since LiDAR is only supported by the newer iPhone models, in all our experiments, we use the iPhone 12 Pro as the LiDAR transmitter. Since the iPhone 12 Pro was a new expensive device, we could not remove the filter on the camera without trial and error risking breaking the camera and voiding warranty. Instead we use an older Samsung Galaxy Note 10+ as a receiver and co-locate its filter-removed  camera right next to the iPhone 12 Pro's camera. As mentioned earlier, iPhone manufacturers have access to the raw images and hence in practice, they can incorporate these techniques using a single  smartphone without the need for physically altering cameras. After the laser speckle pattern is recorded, the blood slide is left in the open at room temperature for 10 minutes. When blood is exposed to air in this time several chemical processes occur in the blood including the formation of fibrin-platelet clots which influence the viscosity of the blood samples and eventually coagulated. The sample is then placed in front of the smartphone setup and the laser speckle pattern is recorded. This procedure was followed for all 30 blood samples. We note that the clotting process can be accelerated to within 10 to 14 seconds by adding thromboplastin tissue factor as an activator to trigger coagulation along the extrinsic pathway of the coagulation cascade.

We plot the correlation curves for the 30 non-coagulated and coagulated whole blood samples in Fig.~\ref{fig:blood}. The shaded gray and red regions denote the various curves we extract for each of these two classes. The plots show that the correlation curves for the non-coagulated samples decorrelate quickly owing to the changing laser speckle pattern in the blood. The curves also converge to a smaller terminal value indicating a lower frame correlation coefficient. This is because the number of pixels that have changed intensity values between frames, and the difference between pixel intensity values is comparatively large. Additionally, there is a wide spread in the correlation curves of the non-coagulated blood samples which is due to natural variations in blood thicknesses and variation in the amount of time each batch of samples was preserved in the refrigerator for, which in turn can affect its viscosity. 

In contrast, the correlation curves for the coagulated sample are much flatter. This is because when the blood has coagulated there is little to no dynamic speckle movement. Instead what is seen is a mostly static speckle pattern being scattered from the coagulated blood. Because of this, there are few pixels that have changed values, and so the terminal value of the frame correlation coefficient is higher. We can also observe that there is less variance in the correlation curves generated for these clotted samples.

Next, we visualize the spread in normalized viscosity coefficients for each these two blood sample classes in Fig.~\ref{fig:blood}. These coefficients are computed using the algorithm described in~\xref{sec:processing}. The plots show  a similar distribution in thee spread for each of the two blood sample classes and a clear distinction between coagulated and non-coagulated blood. The variance in the curves is because of natural variations in viscosity seen in practice in blood samples across humans.

We then run univariate analysis on the correlation curves from these two classes and show the most predictive features in the curve for making a classification.  Fig.~\ref{fig:blood}  shows that the most predictive values are the correlation values between the first frame and the frames immediately after it. Note that these results show the ability to distinguish between coagulated and uncoagulated blood. This is the necessary step required to compute the Prothrombin time (PT) that measures the coagulation time in clinics. We expect subsequent clinical studies are required to compare the PT time computed with the smartphone LiDAR system with  clinical-grade coagulation analysers. 

Finally, we train a classifier on black and white images of the laser speckle fluctuations for non-coagulated and coagulated blood samples. Specifically, we train a support vector machine with an RBF kernel on images like those in Fig.~\ref{fig:fig2}. These images  do not contain any color data, and only contain the speckle information with viscosity information. Our \textit{training dataset} consists of 720 images from 18 coagulated and non-coagulated samples. A \textit{separate, held-out test set} consisting of 240 images is obtained from 6 separate coagulated and non-coagulated samples. The accuracy of our classifier is 98.8\% (Fig.~\ref{fig:blood}d).

\subsection{Milk Sensing}
\label{sec:milkresults}
The second application we evaluate is to use the smartphone LiDAR to perform sensing of milk, in particular, tracking fat contents and identifying adulterated milk.

\vskip 0.05in\noindent{\bf Tracking fat content in milk.} Being able to track the fat content of milk through an objective measure of viscosity is a useful measure in the quality control of food production. Traditional measures of measuring fat content through the Babcock test and Gerber method require large quantities of milk or chemical reagents. Being able to ensure the controlled quality of fat content is important in ensuring that milk samples have not been tampered with due to changes in chemical composition.

In this study, we aimed to investigate if our sensing system could be used to track fat content in milk. We identified five different types of milk products with varying content. Specifically we selected skim milk with 0\% fat, 1\%, 2\% fat milk, whole milk with 3.25\% fat and cream. Our hypothesis is that our system would be able to compute viscosity coefficients for each of these milk types that matches their relative viscosity ordering. In addition, the goal is that our system produces viscosity coefficients that correlate with ground truth measures of viscosity.

\begin{figure}[t]
  \centering
    \begin{subfigure}[b]{0.49\textwidth}
    \includegraphics[width=.32\textwidth]{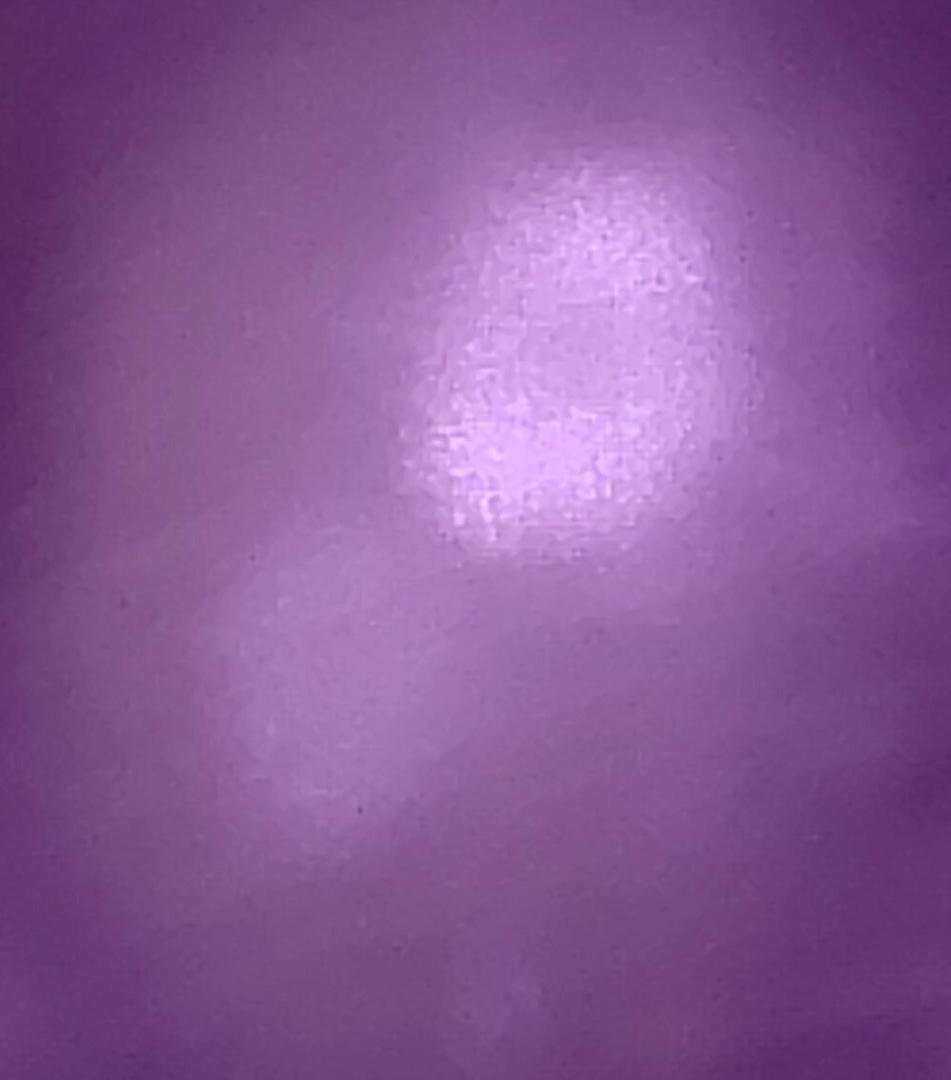}
    \includegraphics[width=.32\textwidth]{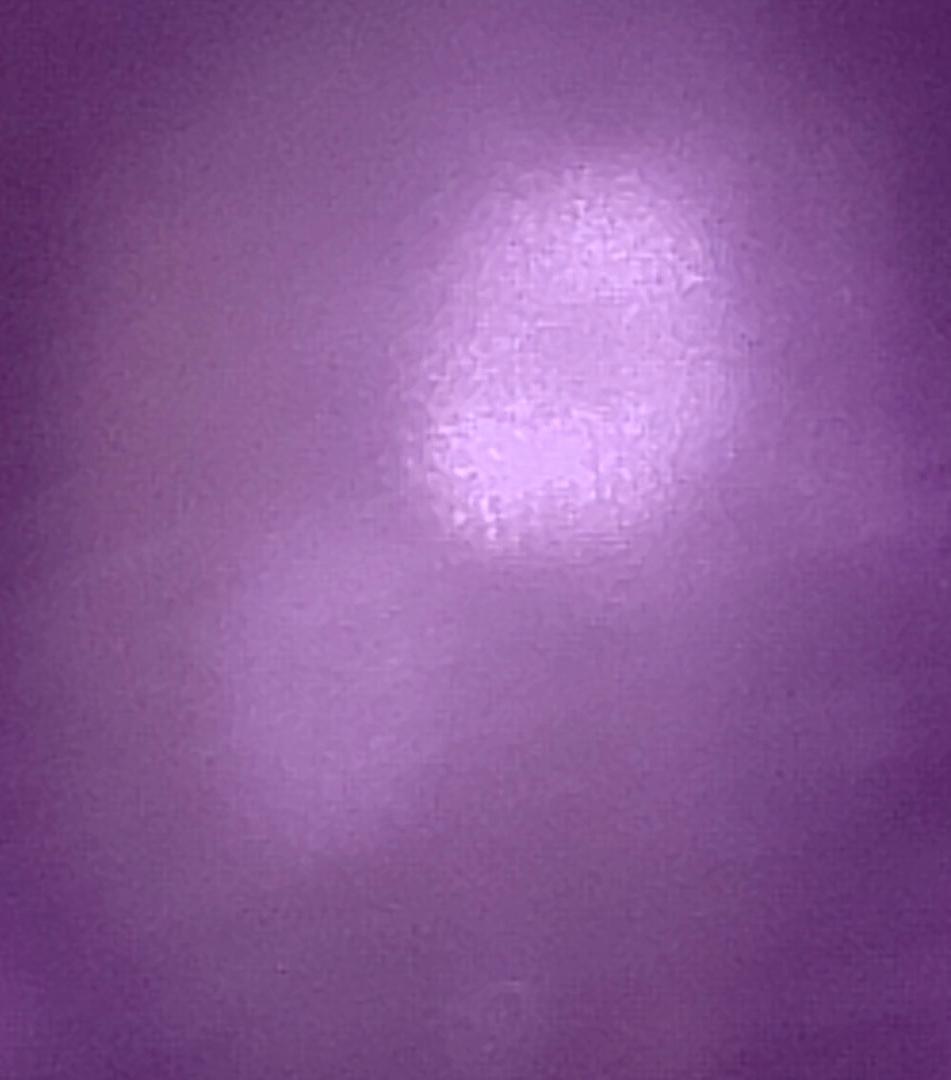}
    \includegraphics[width=.32\textwidth]{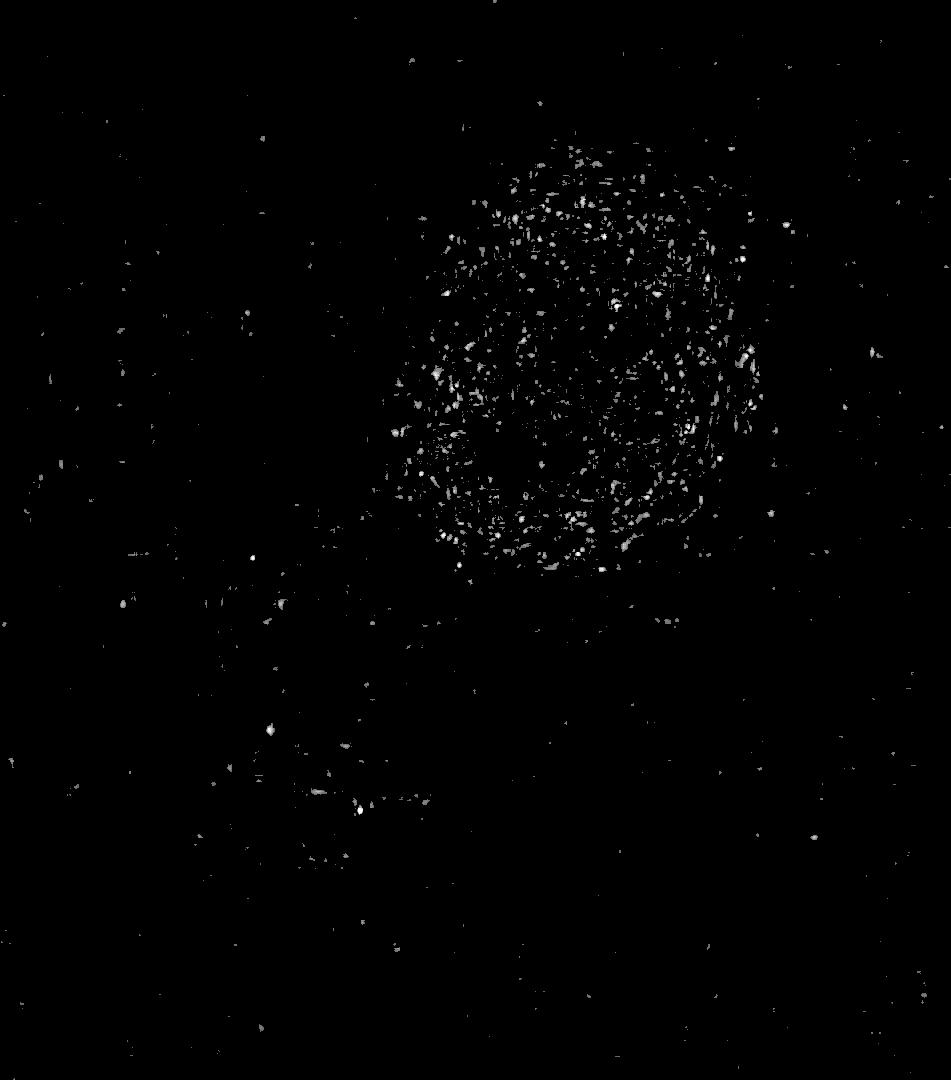}
    \caption{Skim milk (0\% fat)}
    \end{subfigure}
    \begin{subfigure}[b]{0.49\textwidth}
    \includegraphics[width=.32\textwidth]{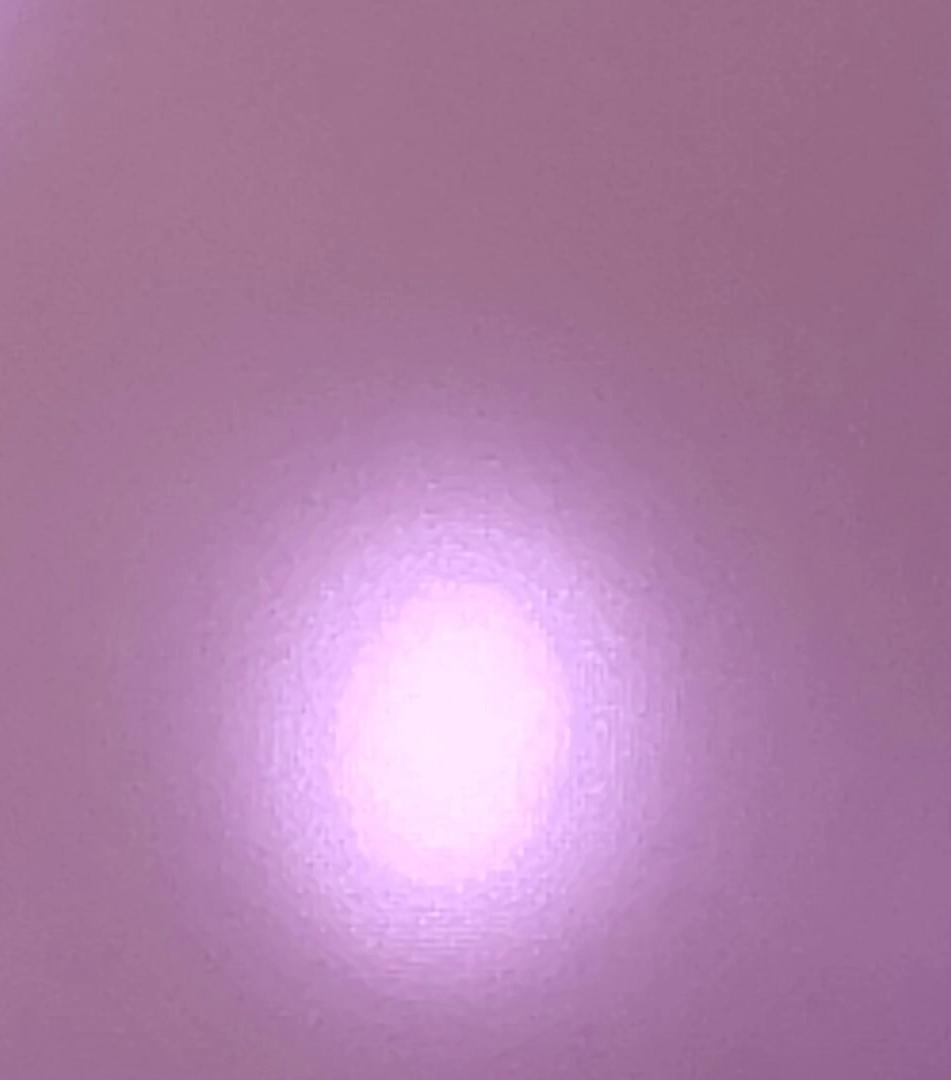}
    \includegraphics[width=.32\textwidth]{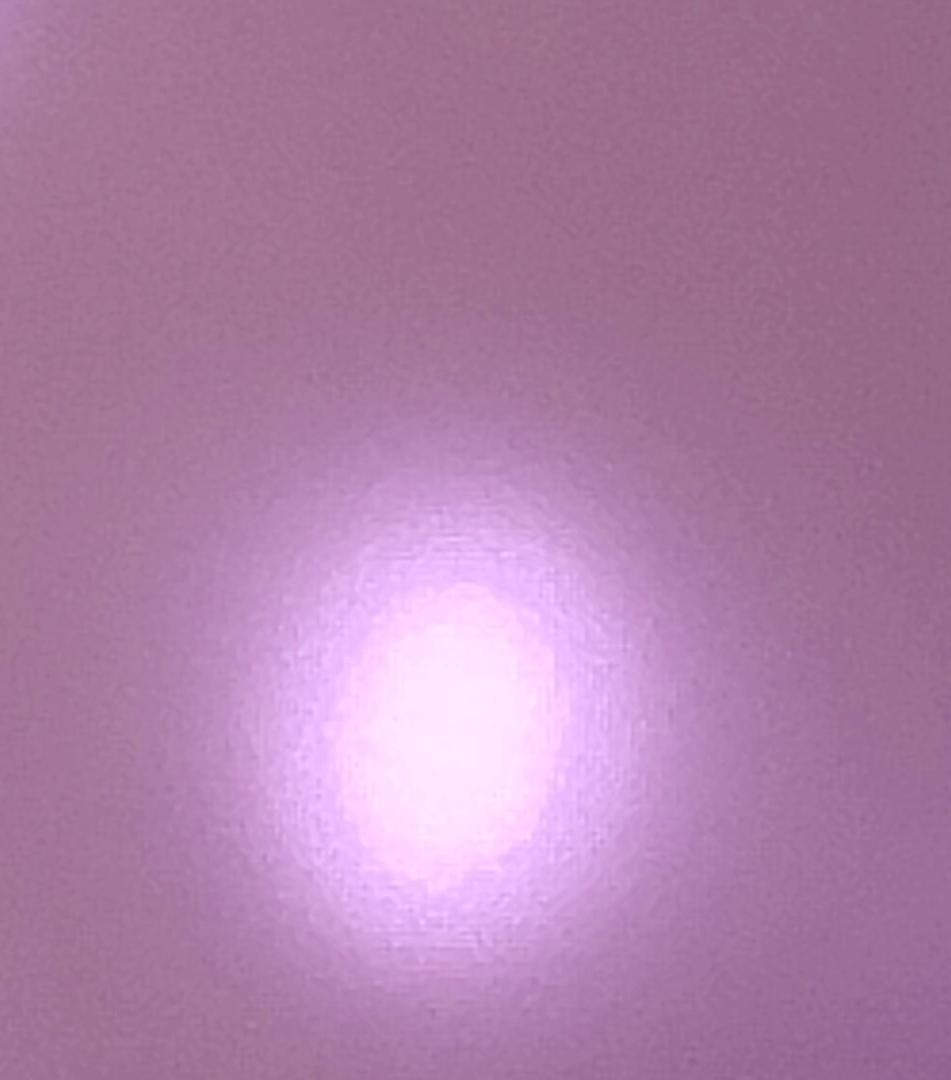}
    \includegraphics[width=.32\textwidth]{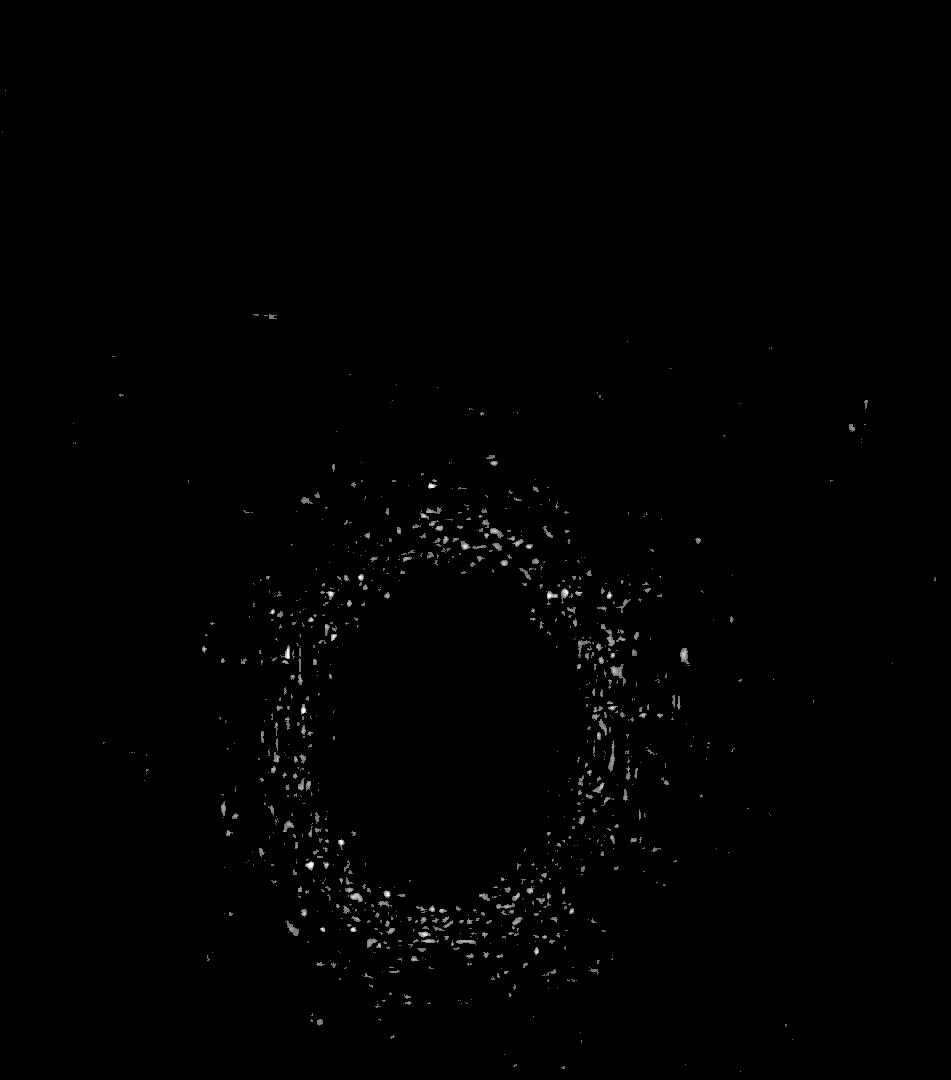}
    \caption{Heavy whipping cream (38\% fat)}
    \end{subfigure}
\vspace{-0.15in}
\caption{Example laser speckle patterns at frame $i$ (left) and frame $i+9$ (middle) for two different types of milks. The image on the right shows the difference between the two frames and highlights the pixels which have fluctuated across this time period.}
\label{fig:examples}
% \vskip -0.15in
\end{figure}

\begin{figure}[t]
  \centering
    \includegraphics[width=\textwidth]{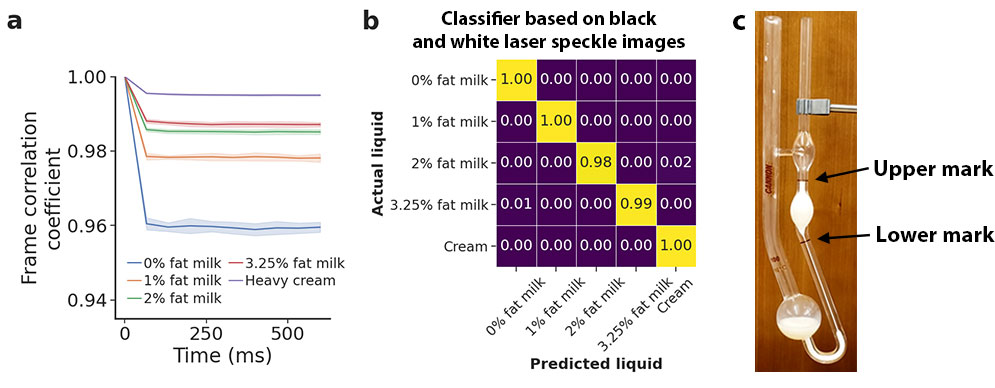}
    
    \includegraphics[width=.32\textwidth]{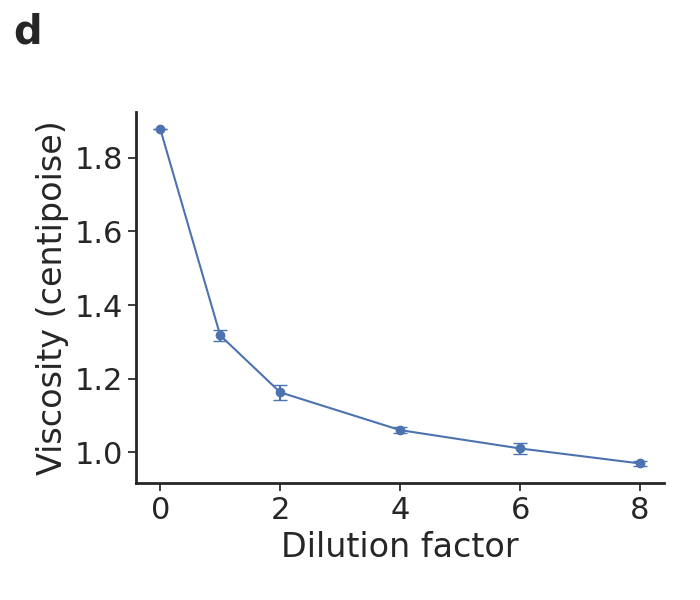}
    \includegraphics[width=.32\textwidth]{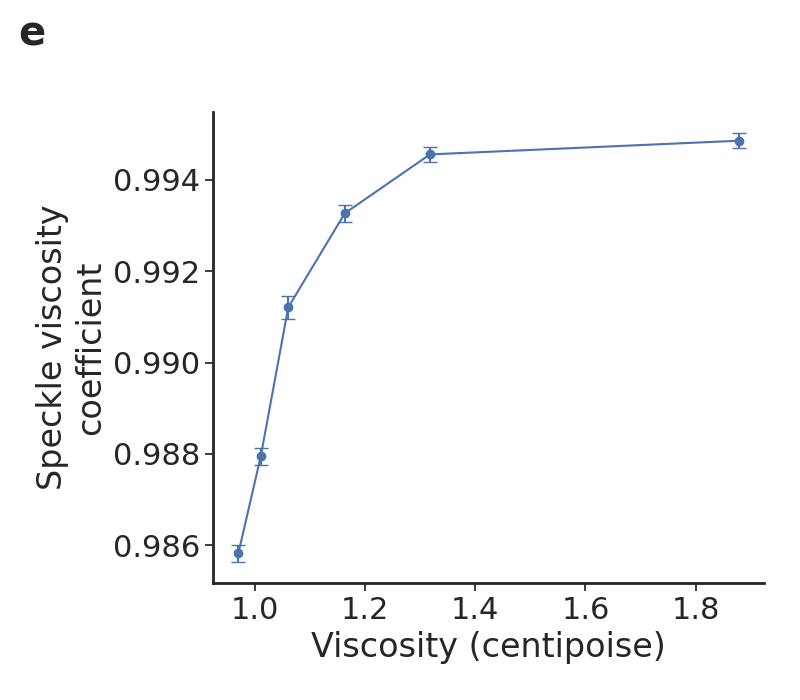}
    \includegraphics[width=.32\textwidth]{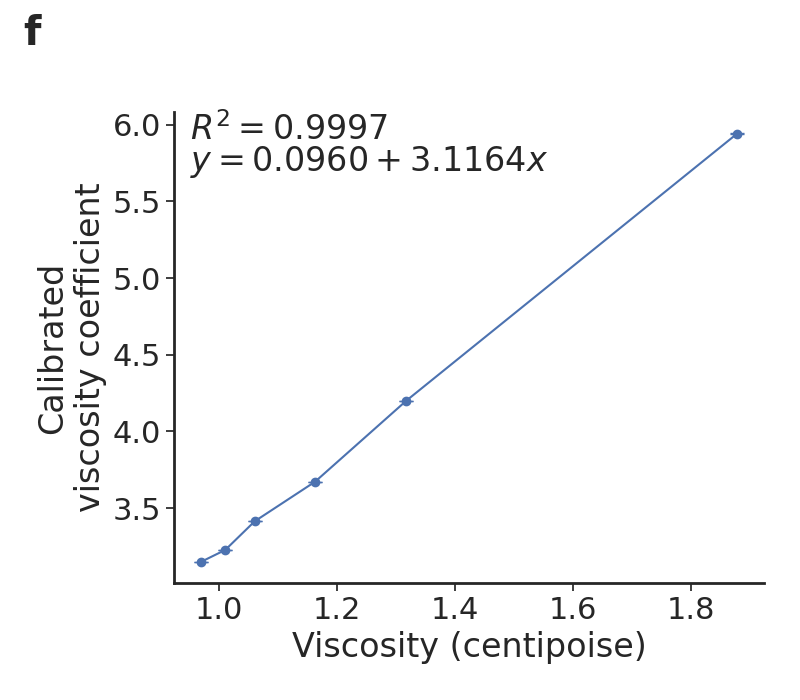}
\vspace{-0.1in}

\caption{Sensing milk fat content. (a) Shows the correlation curves for milks of different fat contents. (b) Confusion matrix showing classification accuracy at distinguishing between milks of different fat contents. (c) Ostwald viscometer used to obtain ground truth measurements of viscosity. (d) The ground truth viscosity measured for different dilutions of whole milk. (e) The relationship between ground truth viscosity and the speckle viscosity measurement. (f) The correlation between ground truth viscosity measurements and the calibrated viscosity coefficient.}
% \vskip -0.2in
\label{fig:fat}
\end{figure}

Fig.~\ref{fig:examples} shows the speckle pattern obtained for a sample of skim milk and cream at the $i^{th}$ frame and the $i+9^{th}$ frame. We plot the normalized pixel differences between the two frames to indicate the pixel intensity changes as a result of speckle pattern fluctuations. We can see there is a lot of movement throughout the speckle pattern of skim milk, while there is comparatively less movement in terms of intensity magnitude and in terms of number of pixels fluctuating. Due to the larger number of fluctuating pixels in skim milk, the terminal value of the frame correlation  coefficient will also be lower compared to cream.

For this study, we pipette 20 \textmu l of each milk solution onto a glass slide and place the slide in front of the smartphone LiDAR so the speckle image can be recorded. This is repeated for all the different types of milk for a total of three separate samples per each milk type.

Our results in Fig.~\ref{fig:fat}a show the correlation curves obtained for the five different types of milk types. The plots show that the viscosity coefficients and terminal frame correlation coefficients obtained by the receivers match the expected order of viscosity. Less viscous milks, particularly skim milk with 0\% fat, exhibit more measurement variance than the most viscous milk types like cream. 

We investigate if we can classify between these five different milk types. We train a support vector machine with an RBF kernel to classify between the black and white images of speckle fluctuations for each of these milks, like those in Fig.~\ref{fig:examples}. These images {do not} contain any color data, and our classifier only leverages the speckle pattern encoding viscosity information. Our \textit{training dataset} consists of 265 images, with 53 images for each of the milk classes. Each set of 53 images is obtained from a single video recording of the liquid. A \textit{separate, held-out test set} consisting of 795 images is obtained from three separate volumes of the liquid. In other words, there is no overlap between the training and test sets. The average accuracy of our classifier across these three measurement sessions is 99.5 $\pm$ 0.2\% (Fig.~\ref{fig:fat}b).

To determine if our system produces viscosity coefficients that correlate with ground truth measures of viscosity, we record the speckle images and ground truth viscosity for different dilutions of whole milk. Specifically, the whole milk is diluted with water at increasing dilution ratios from 0 to 8. Ground truth measurements of viscosity were obtained using an Ostwald viscometer (Fig.~\ref{fig:fat}c). The viscosity $\eta$ of a fluid is calculated by first measuring the amount of time $t$ for the liquid to fall between the upper and lower mark of the viscometer. The following equation is then used to calculate the viscosity of the liquid:
$$ \eta = \frac{\rho t}{\rho_{ref} t_{ref}} \cdot \eta_{ref}$$

Here, $\rho_{ref}=0.997~g/ml$ and $\eta_{ref}=0.8937~cP$ corresponds to the density and viscosity of a reference fluid, specifically water, and $t_{ref}$ corresponds to the time for the reference fluid to fall between the upper and lower marks of the viscometer. Fig.~\ref{fig:fat}d shows the ground truth viscosity measurements for different dilutions of whole milk.

Fig.~\ref{fig:fat}e shows the non-linear relationship between the ground truth viscosity measurements and the speckle viscosity coefficient measured on the smartphone. We note that the speckle viscosity coefficient is only a proxy for viscosity, as obtaining a closed-form equation to calculate viscosity depends on other properties of the liquid such as the size of the light-scattering particles~\cite{hajjarian2015optical,hajjarian2015estimation,hajjarian2020tutorial}. In order to obtain a mapping from the speckle viscosity coefficient to the ground truth measurement, a one-time calibration procedure needs to be performed for each class of liquids.

We perform the calibration procedure for different dilutions of whole milk by fitting the measured viscosity coefficients for different dilutions to a cubic curve. To evaluate this calibration procedure, we then perform three separate measurements on new volumes of milk dilutions. The coefficients of the cubic curve are then used to normalize these new measurements. Fig.~\ref{fig:fat}f shows the linear relationship between the calibrated viscosity measurements and ground truth viscosity values.

\begin{table}

\begin{center}
\caption{Milk adulterant solutions used in this study. Only 20 \textmu l of the mixture is used for testing.}
 \begin{tabular}{| c | c |}\hline
 {\bf Milk solution} & {\bf Recipe} \\ \hline
 Milk & Whole milk, 3.25\% fat \\  \hline
 Milk and detergent & 2ml in 20ml milk (9\% concentration) \\ \hline
 Milk and salt & 1.25g salt in 20ml of milk\\ \hline
 Milk and cornstarch & 1.25g cornstarch in 20ml of milk \\ \hline
 Milk and water & 2ml in 20ml milk (9\% concentration) \\ \hline
 Milk and xanthan gum & 1.25g xanthan gum in 20ml of milk\\ \hline
\end{tabular}
\label{table:mtable}
\end{center}

% \vskip -0.2in
\end{table}

\vskip 0.05in\noindent{\bf Identifying adulterated milk.}  Milk is one of the most commonly consumed beverages globally. However, there have historically been major scandals of significant amounts of milk supply which have been adulterated~\cite{milkscandal}.  Being able to track the adulteration status of milk in a contactless manner with an accessible device like a smartphone instead of using chemical reagents is important for creating an accessible screening tool given the widespread consumption of milk.

Our goal here is to study if our system is able to differentiate between unadulterated whole milk and samples of whole milk that have been adulterated with different substances. To do this, we  identified a list of common household substances that could end up inadvertently contaminating milk. Table~1 shows the list of different milk adulterants we identified. It should be noted that these adulterants notably alter the viscosity  of the substance, which would be reflected in a change in the laser speckle pattern. For this experiment, we first prepare different contaminations of milk, and pipette 20~\textmu l of each solution onto a glass slide. Each sample is then placed in view of our smartphone LiDAR sensing setup and the laser speckle pattern is recorded for the control sample and each of the adulterant samples. This is repeated two more times, so each milk solution is imaged by our system for a total of three times. All the samples were left for a while at room temperature before running these experiments.

Fig.~\ref{fig:adult}a shows the correlation curves obtained for each of the milk solutions. The curves show that the laser speckle pattern in whole milk sample decorrelates more quickly than with the adulterated samples. In other words, the adulterated samples appear more viscous than the whole milk sample. When the adulterant is cornstarch and xanthan gum, the sample appears to be notably more viscous based on how it does not move as freely along the glass slide. In the case of water, detergent and salt, the sample becomes less viscous due to what one would expect to be dilution of the milk's fat content. However, these adulterants also have a negative effect on the visibility of the speckle pattern, making it more challenging to track the dynamic changes in speckle movement. In particular, the water and detergent decreases the opacity of the solution making it slightly more transparent. This reduces the amount of LiDAR transmissions scattered from the liquid sample. Similarly the speckle from the salt dilution appears less visible and changes less. Fig.~\ref{fig:adult}b shows the spread in viscosity coefficients for the sample of control milk and adulterated milk, illustrating that they do not overlap and can be easily differentiated.

\begin{figure}[t]
\centering
\includegraphics[width=.48\textwidth]{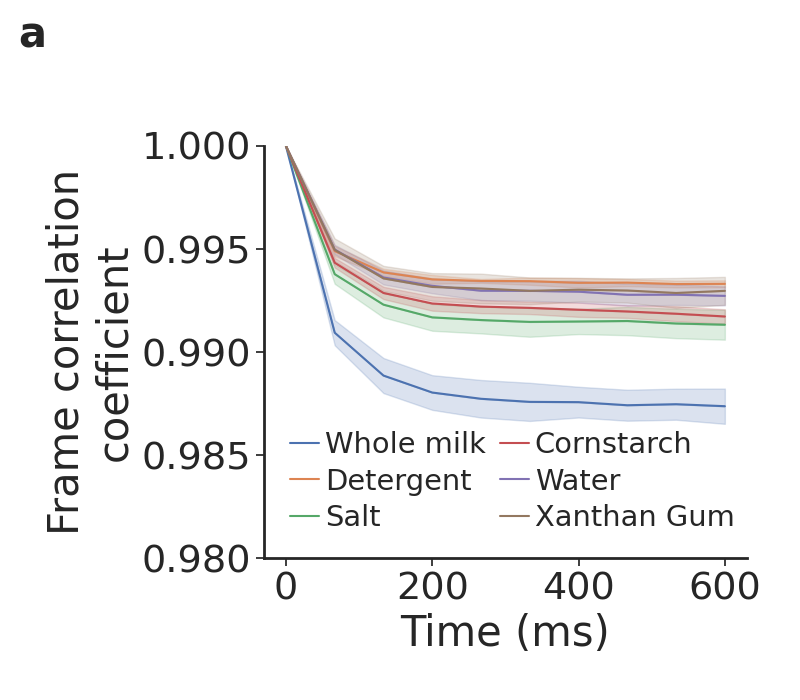}
\includegraphics[width=.48\textwidth]{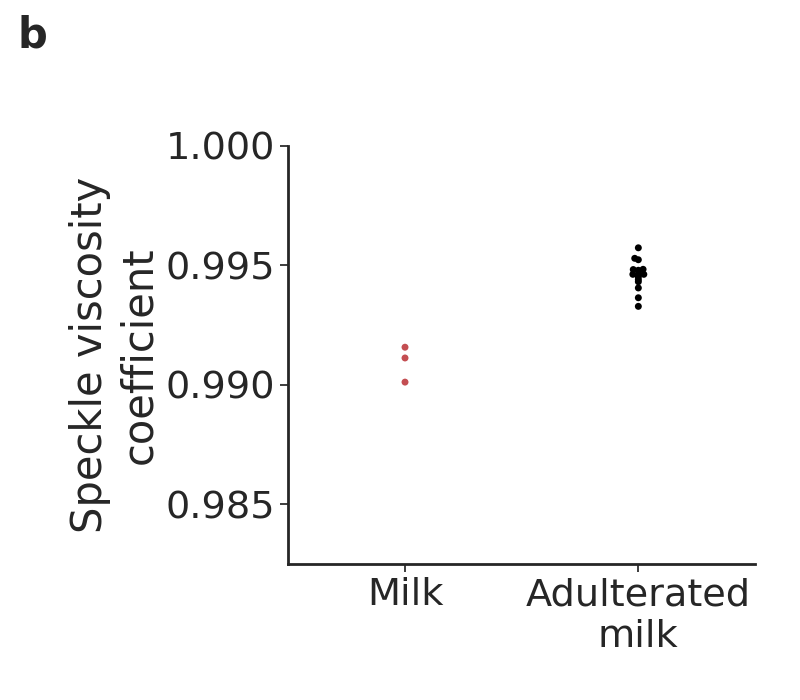}
\vspace{-0.1in}
\caption{Detecting adulterated milk. (a) Correlation curves show that unadulterated milk has a viscosity coefficient that is distinct from the adulterated recipes. (b) Spread of normalized viscosity coefficients for unadulterated and adulterated milk recipes.}
\label{fig:adult}
% \vskip -0.2in
\end{figure}

\subsection{Laser Speckle-based Liquid Classification}
\label{sec:liquidresults}

\begin{figure*}[t!]
\includegraphics[width=\textwidth]{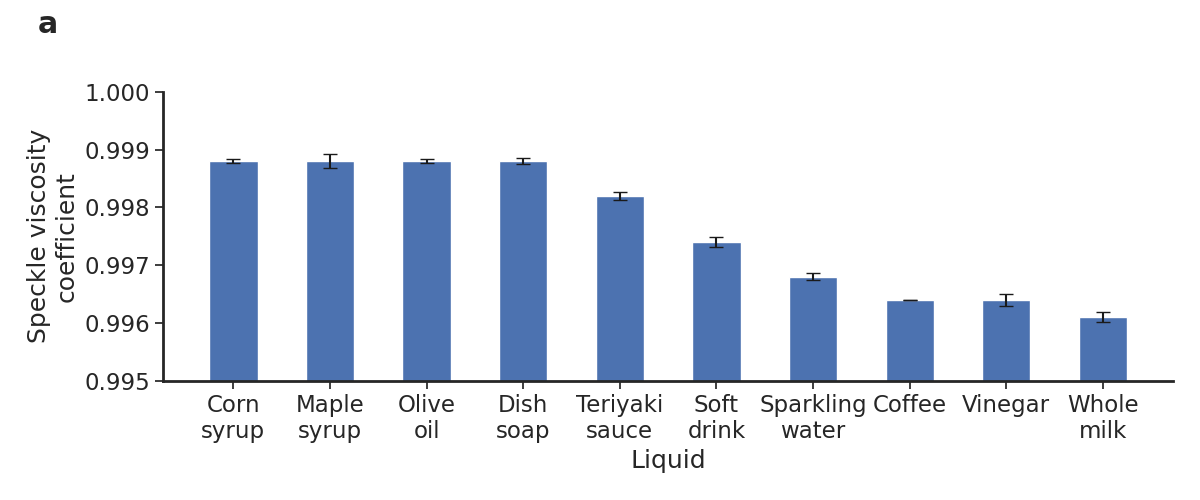}
\includegraphics[width=\textwidth]{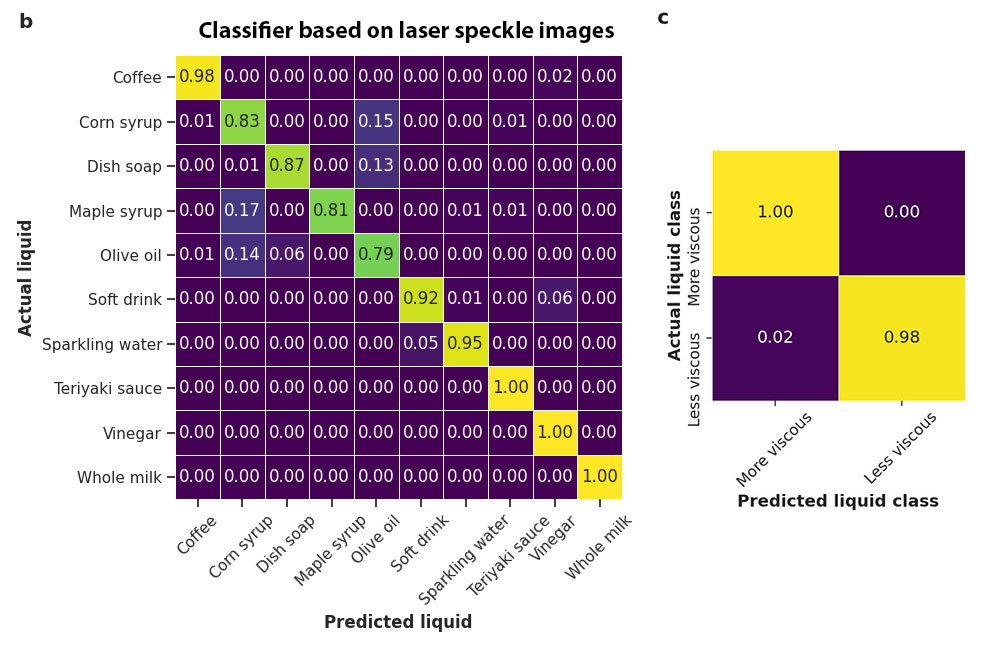}
\vspace{-0.2in}
\caption{Classifying between different liquids. (a) Viscosity coefficient calculated for each of the different liquids. (b) Confusion matrix showing classification accuracy across ten different liquids of varying viscosity values. (c) Confusion matrix showing binary classification accuracy at distinguishing between viscous and less viscous fluids. }
\label{fig:multiliquid}
\end{figure*}

Finally, we evaluate if our smartphone LiDAR system is able sense the viscosity differences across a broader range of liquids with varying viscosities levels, and classify between them.

In this study, we identified ten different liquids of varying viscosities including thick viscous liquids like corn syrup, maple syrup, olive oil and dish soap, to less viscous liquids like sparkling water, coffee, and vinegar. We hypothesize that our system would be able to compute LiDAR viscosity coefficients that correlates with their relative viscosity ordering. Additionally, we investigate if we can classify between these different liquid classes using images of speckle fluctuations.

Our result in Fig.~\ref{fig:multiliquid}(a) shows the ten liquid classes ordered by the LiDAR viscosity coefficients calculated by our system. The plot shows that more viscous liquids such as corn syrup, maple syrup, olive oil and dish soap all have a high viscosity coefficient, and are also relatively similar to each other. This is because the speckle pattern in these liquids do not change much, and there is little speckle movement across these liquids. Less viscous fluids including a soft drink, sparkling water, coffee, vinegar and whole milk have lower viscosity coefficients.

To determine if we are able to distinguish between the ten liquids based on the speckle pattern, we train a support vector machine with an RBF kernel to classify between image frames of the speckle fluctuations obtained from each of liquids; note that these frames are the black and white speckle patterns shown in Fig.~\ref{fig:fig2} and {\bf do not} include any color data. This makes sure that our classifier is using only the speckle pattern that encodes the viscosity information. Specifically, our dataset is comprised of  differences between image frames of the speckle pattern like those in Fig.~\ref{fig:examples}, which highlight laser speckle fluctuation. Our \textit{training dataset} consists of of 530 images, with 53 images for each of the ten liquid classes. Each set of 53 frames were obtained from the same video recording of the liquid. Our \textit{separate, held-out test set} consists of 1590 image frames, obtained from three fresh volumes of each of the liquids, i.e., there is no overlap between the training and testing data and the test data set is not even the same measurement as the training data. Across these three measurement sessions, the accuracy of our classifier is 91.5 $\pm$ 3.9\%. Fig.~\ref{fig:multiliquid}b shows the confusion matrix of the classifier. The most common misclassifications were between viscous classes, specifically corn syrup, maple syrup and olive oil. These liquids also have similar viscosity coefficients.

Next, we group these liquids into two classes: more viscous and less viscous. The syrups, oil and soap are classified as more viscous, and the remaining liquids are classified as less viscous. When our classifier is retrained on these binary labels, the average accuracy across the three measurements sessions is 99.5 $\pm$ 0.2\%. The confusion matrix of this classifier (Fig.~\ref{fig:examples}c) shows that more viscous liquids are correctly classified 100\% of the time and less viscous liquids are correctly classified 98\% of the time.

\subsection{Benchmark Experiments}
\label{sec:benchmark}

We perform benchmark experiments to understand the effects of various design parameters.

\vskip 0.05in\noindent{\bf Effect of liquid volume.} We first consider the effect of liquid volume on the performance of our system. We captured the speckle pattern for 0\% fat milk, whole milk and cream and varied the sample volume to be 10, 30 and 50 \textmu l. We compute the normalized viscosity coefficients as described in~\xref{sec:processing} and plot them in Fig.~\ref{fig:shutter}a. The samples are placed at a distance of 5~cm and the measurement with each type of liquid was repeated three times. We find that the order of viscosity coefficient remains preserved across these volumes and the speckle pattern looked similar across liquid volumes. The laser speckle was able to cover the full surface area of both the 10 and 50~\textmu l sample, so increasing the volume amount had no significant effect on viscosity index. We however  note that using larger samples volumes such as 2~ml of milk in a petri dish would have an effect on how the speckle pattern looked. This is because such a volume in a petri dish would increase the effective opacity of some fluids. Specifically, 0\% fat milk is slightly translucent and the speckle from the surface the sample is resting on is visible in the captured video. But if 2~ml of 0\% fat milk were in a petri dish it would appear opaque and the static speckle from beneath the petri dish would not be visible.

\begin{figure}[t]
  \centering
    \includegraphics[width=.32\textwidth]{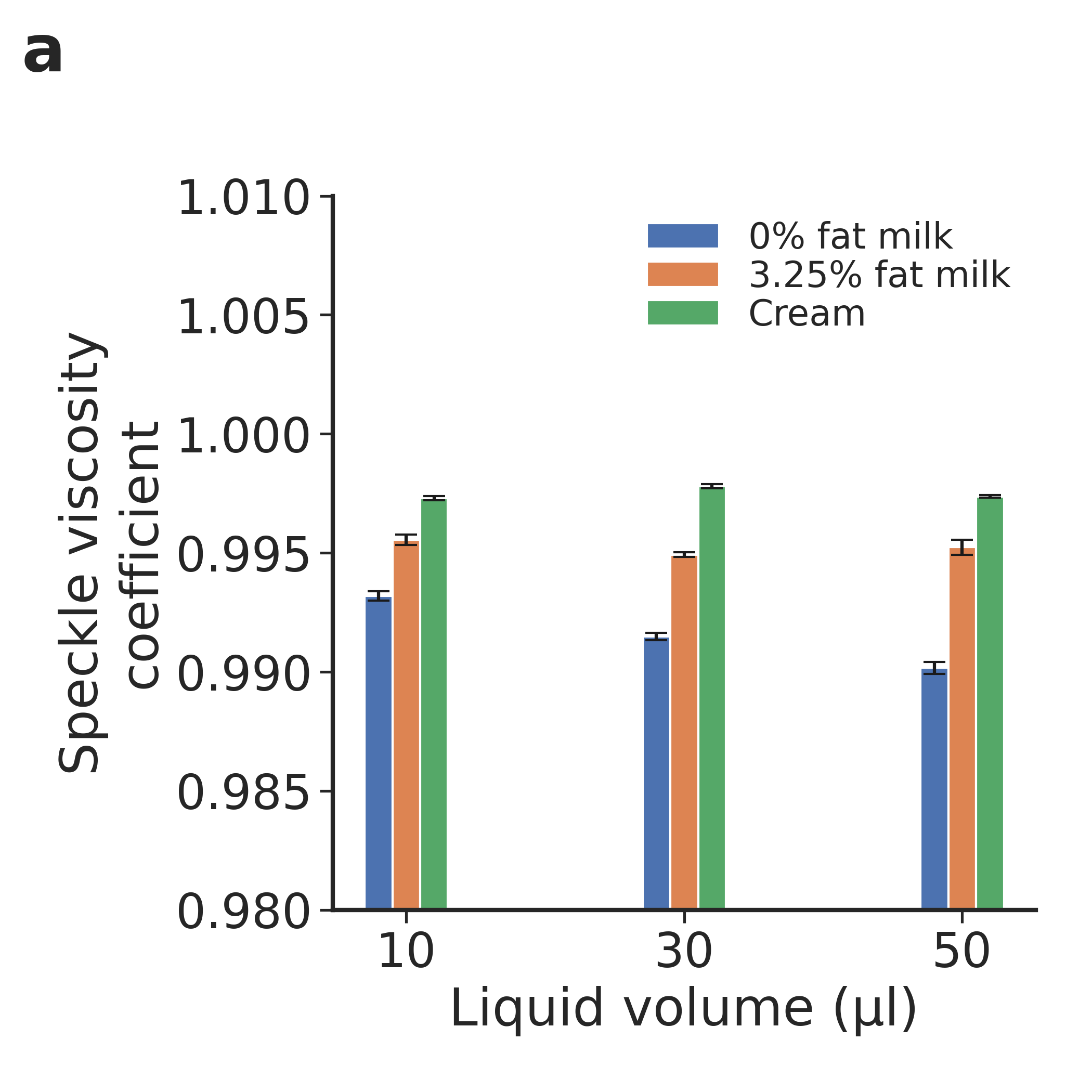}
    \includegraphics[width=.32\textwidth]{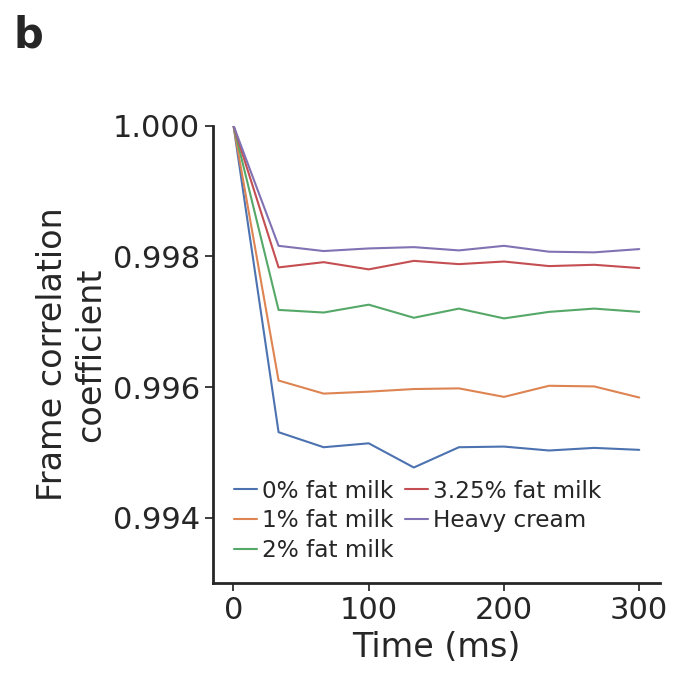}
    \includegraphics[width=.32\textwidth]{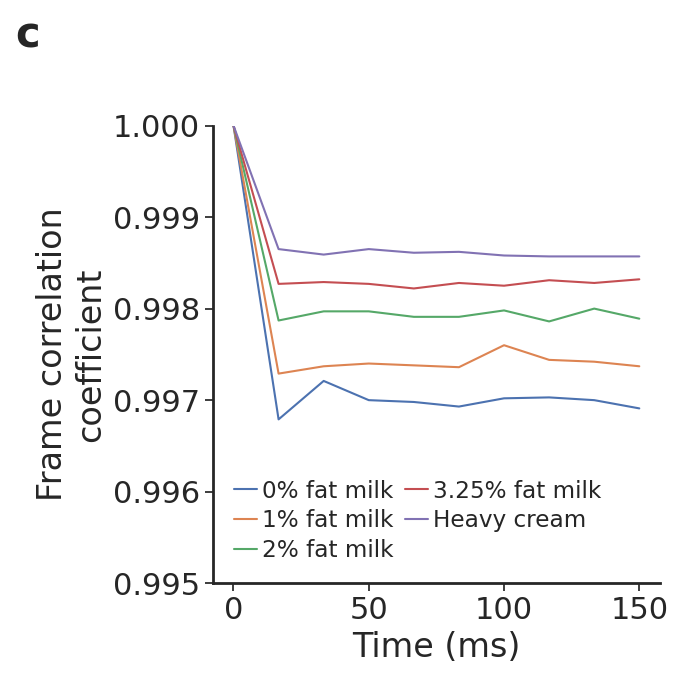}
    \vskip -0.15in
    \caption{(a) Effect of liquid volume on speckle patterns. (b,c) Effect of shutter speed for 1/30~s and 1/60~s captured at 30 frames per second.}
\label{fig:shutter}
% \vspace{-0.2in}
\end{figure}

\begin{figure}[t]
\centering
\includegraphics[width=.32\textwidth]{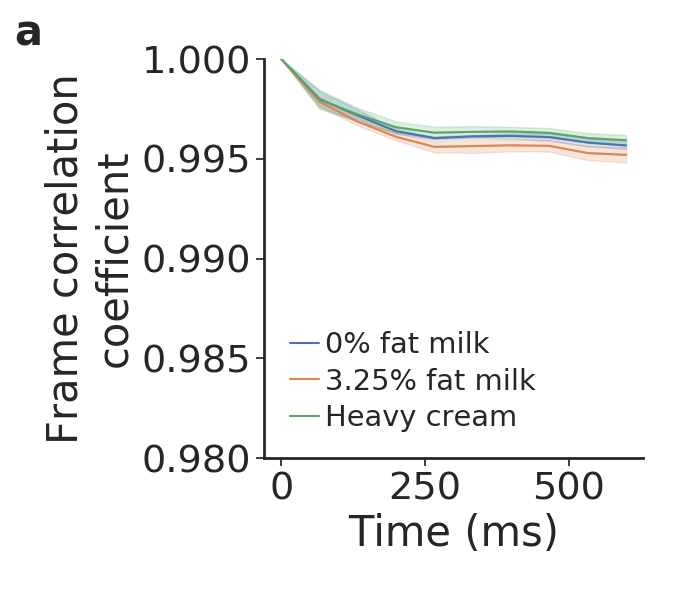}
\includegraphics[width=.32\textwidth]{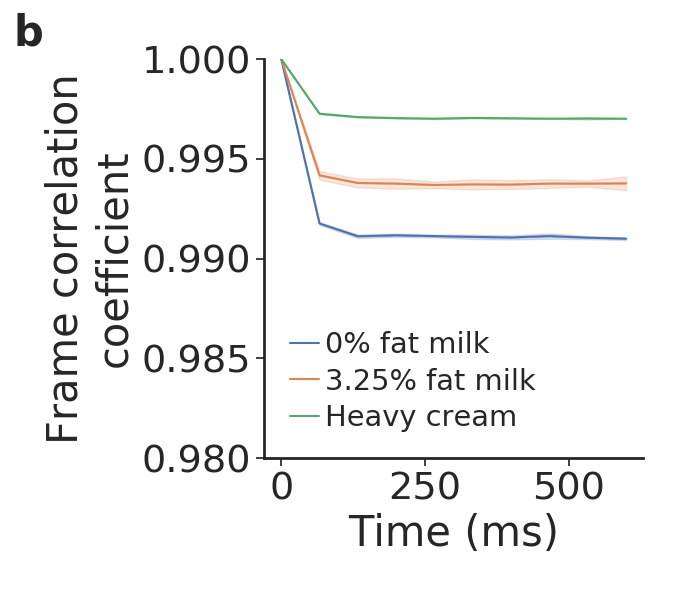}
\vskip -0.15in
\caption{Effect of zoom (2x vs. 8x). A 8x zoom is required to capture the minute speckle variations.}
% \vspace{-0.2in}
\label{fig:zoom}
\end{figure}

\vskip 0.05in\noindent{\bf Effect of shutter speed.} Next we consider the effect of the camera's shutter speed settings  on our ability to capture the laser speckle patterns. Here we consider how our algorithm for overcoming the rolling shutter effect works for different shutter speeds. We repeat experiments with 0\% fat milk, whole milk and cream each with a volume of 20~\textmu l. Fig.~\ref{fig:shutter}b,c shows the correlation curves for both shutter speeds of 1/30 and 1/60~s with a frame rate of 30 frames per second.  The plots show that the correlation curves coefficients for different dairy products match the expected viscosity order. We suspect that the spikes in the correlation curves may be a result of some remaining rolling shutter artifact, in particular, the minor wobbling or skewing of pixels. However, it should be noted that these minor artifacts do not have an effect on the overall ordering of the correlation curves.

\vskip 0.05in\noindent{\bf Effect of camera zoom.} Next, we evaluated the effect of two different camera zoom levels. As before we repeat the experiments with 0\% fat milk, whole milk and cream each with a volume of 20~\textmu l and the sample is at a distance of 5~cm. Fig.~\ref{fig:zoom} shows the correlation curves for two zoom settings of 2x and 8x. The plots show that while at a zoom level of 2x, it is possible to still observe the `boiling' and `sparkling' pattern of the speckle moving, it is challenging to differentiate between speckle activity between the samples. As a result, all three samples appear to decorrelate at the same rate. The reason the curves are not completely flat with a correlation value of 1 is because some speckle activity is still visible, and noise in the image  is inevitable. In contrast, at a zoom level of 8x, we can see the correct ordering between the three liquid samples, with a reasonable separation in coefficients between them. This demonstrates that the zoom level is an important design parameter.

\begin{figure}[t]
\centering
\includegraphics[width=\textwidth]{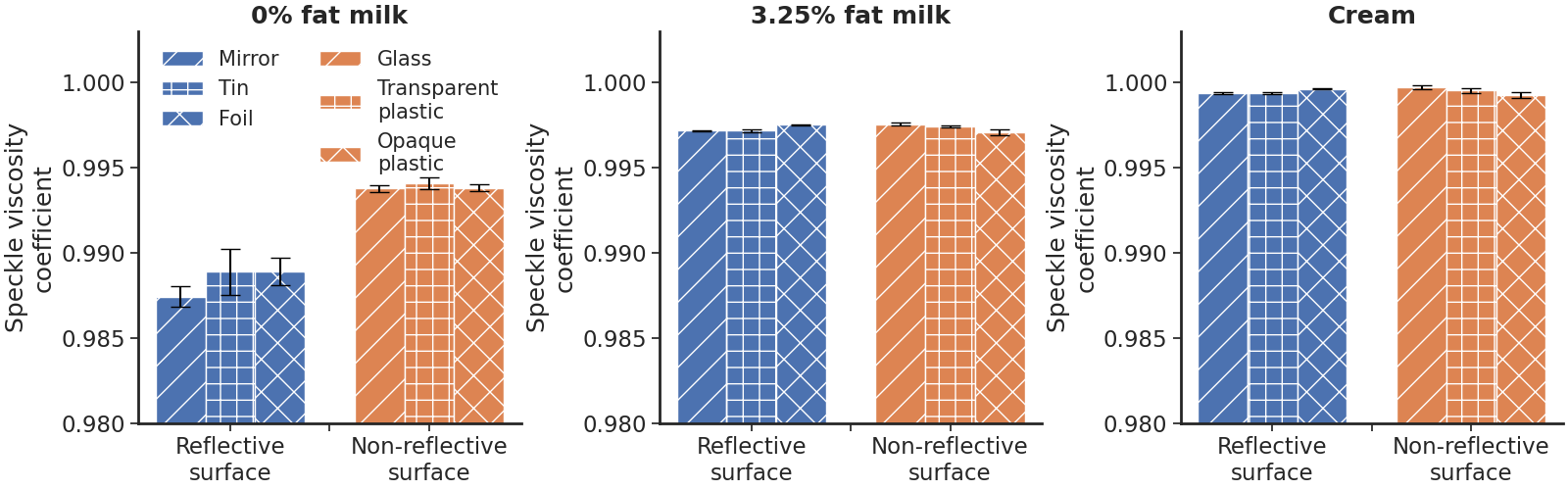}
\caption{Effect of surface material. The speckle pattern requires a non-reflective surface for best performance.}
% \vspace{-0.2in}
\label{fig:ref_mat}
\end{figure}

\vskip 0.05in\noindent{\bf Effect of surface material.} We measure whether the surface material the sample is placed on has an effect on speckle pattern. We select three non-reflective materials including, glass, transparent plastic, and opaque plastic; and three reflective surface materials including a mirror, tin, and foil. We repeat the above experiments with all the other parameters staying constant. Fig.~\ref{fig:ref_mat} shows that the viscosity ordering is preserved across the six different surface materials. Additionally, it shows that the viscosity coefficient for opaque liquids, the 3.25\% fat milk and cream, is consistent across all six non-reflective and reflective surfaces. We observe that the use of a reflective surface material results in a different speckle pattern for the translucent 0\% fat milk, and results in a reduced speckle viscosity coefficient, compared to the non-reflective materials. We observe that the speckle viscosity coefficient is lowest for the glass mirror, which is the most reflective surface used, compared to the tin and foil. The variance in speckle viscosity coefficient is also higher when a reflective surface is used. These results suggest that the system should be used with non-reflective surfaces.

\begin{figure}[t]
\centering
\includegraphics[width=\textwidth]{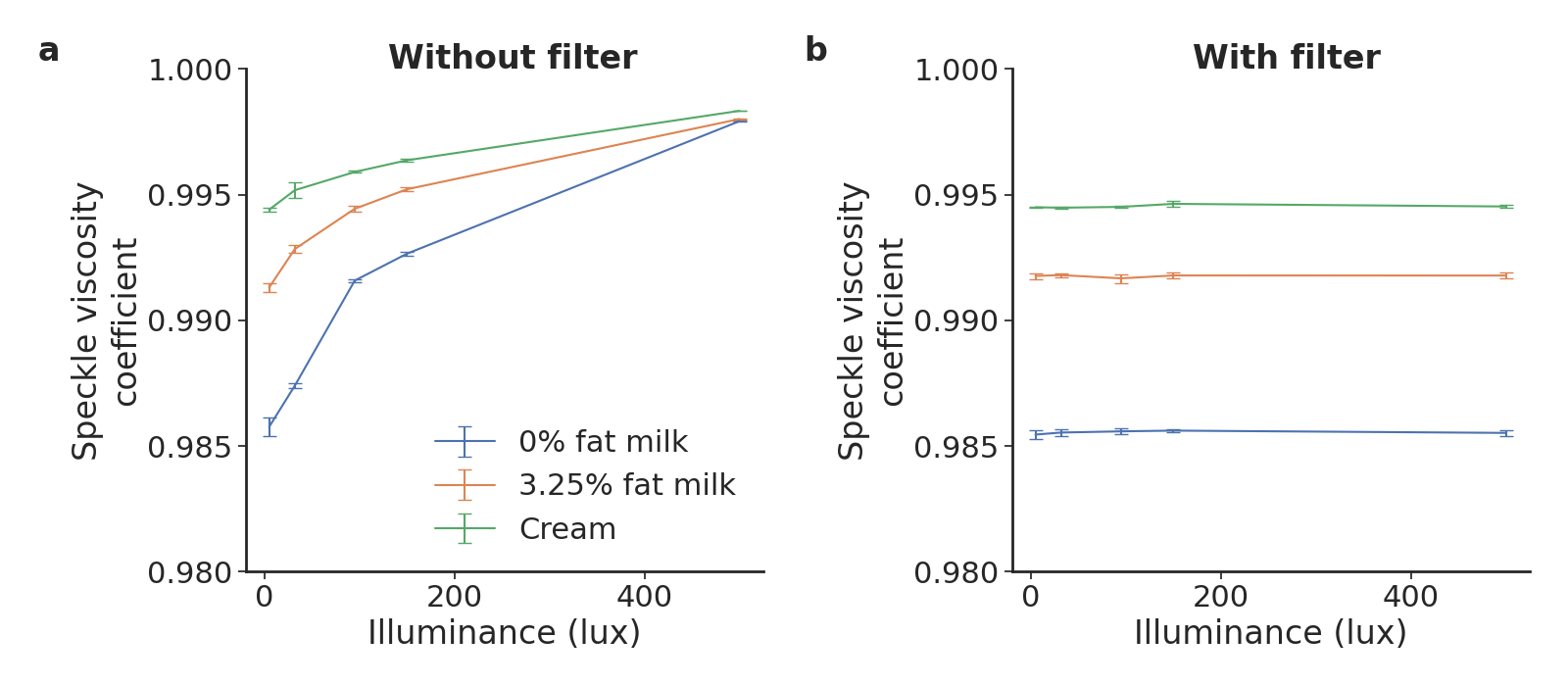}
\caption{Effect of background light intensity. (a) Shows that the speckle pattern requires a low background light for increased visibility of the speckle pattern. (b) Applying a passive filter to filter out visible light preserves visibility of the speckle pattern across different background light intensities.}
% \vspace{-0.2in}
\label{fig:light_filter}
\end{figure}

\vskip 0.05in\noindent{\bf Effect of background light.} We also vary the amount of background light shining on the sample using the flash from a \red{Samsung Galaxy S9} smartphone between 5 and 500 lux. \red{Specifically, the smartphone flashlight was configured to five different brightness levels that had an illuminance level of 5, 32, 95, 150, and 500 lux at the liquid sample. The illuminance level was measured using the Light Meter LM-3000~\cite{lightmeter} iOS app on an iPhone 7, which is calibrated to measure illuminance using the camera on iPhones.} Fig.~\ref{fig:light_filter} shows the results for these experiments at different background light levels. The plot shows that the viscosity ordering is preserved for all illuminance levels, except for the high illuminance level of 500 lux. Additionally, it shows that the speckle viscosity coefficient for all three milks increases non-linearly with illuminance. Increasing background light reduces the differences in speckle viscosity coefficient between liquids, and at high levels, saturates the image making it difficult to discern speckle patterns and distinguish between viscosity ordering. We believe that this is acceptable for liquid testing since it might be possible to control the background light either by placing the smartphone in a darker area or by containing the setup in a closed case to control for background light. When a passive visible light filter (Wratten 87) is applied in front of the smartphone's camera, it can filter this background light, resulting in a consistent speckle viscosity coefficient across all measured illuminance levels for the three milks.

\begin{figure}[t]
\centering
\includegraphics[width=.48\textwidth]{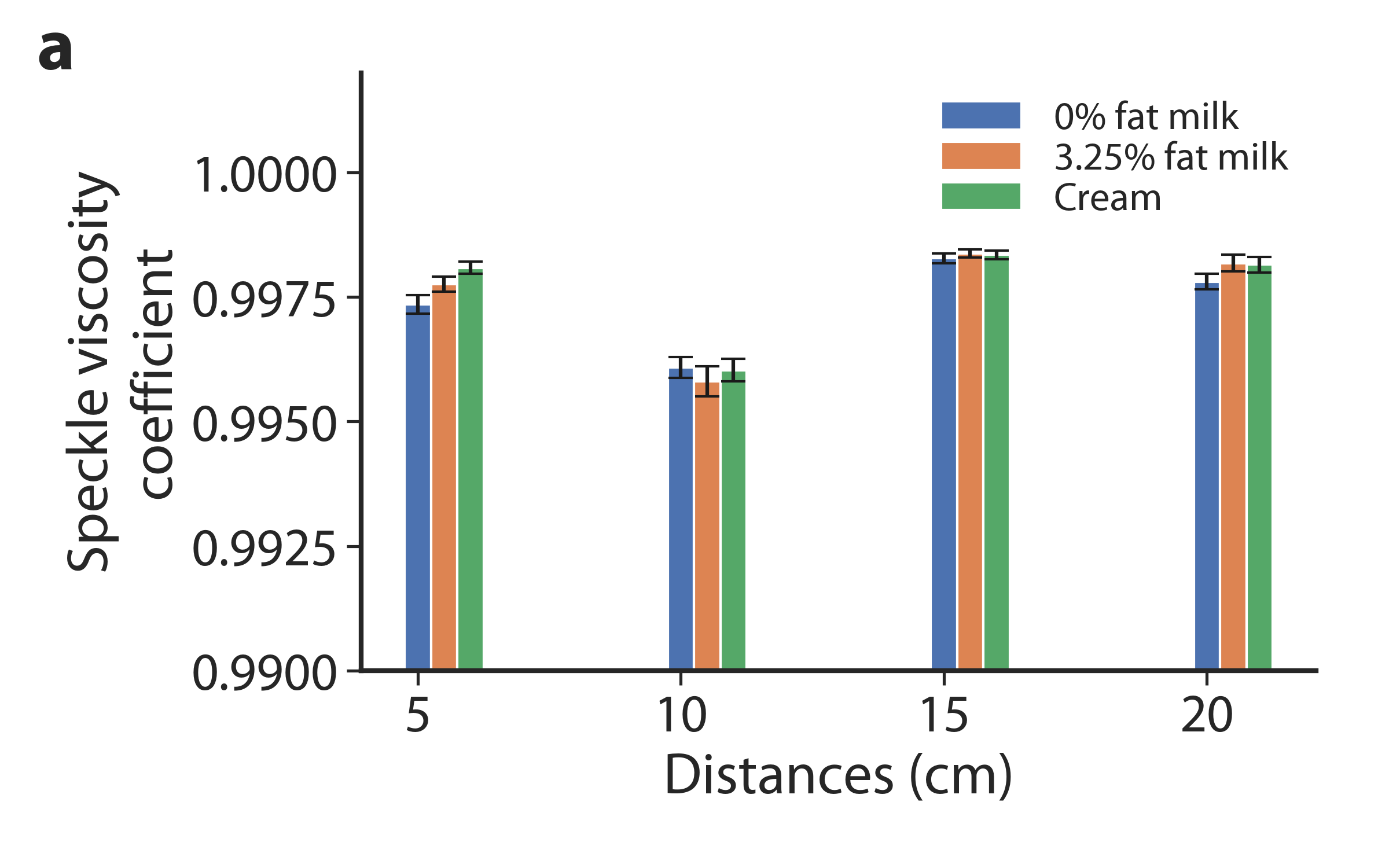}
\includegraphics[width=.48\textwidth]{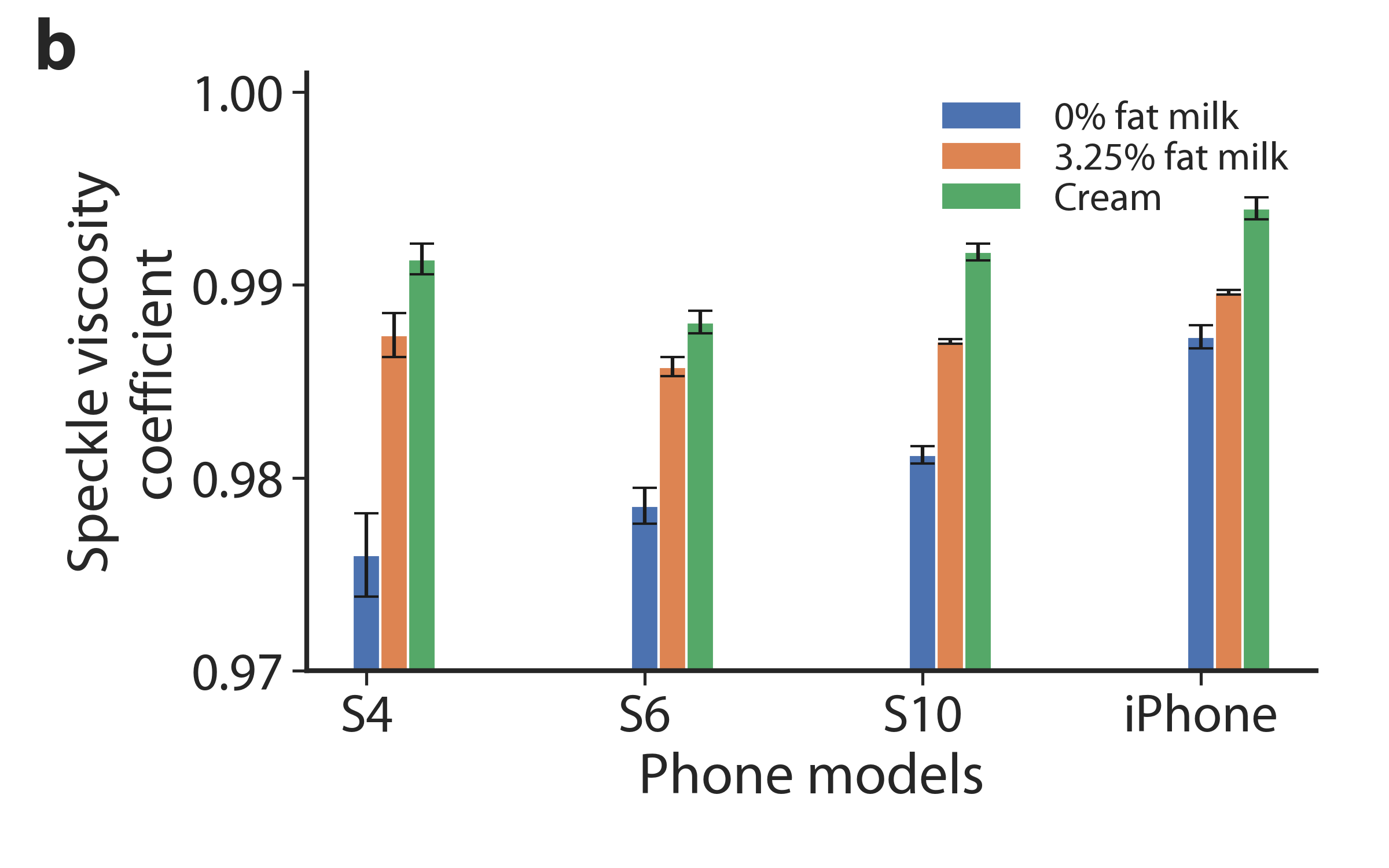}
\caption{Effect of (a) distances and (b) phone camera models. (a) The system requires a close distance of 5 cm for optimal performance. (b) shows that cameras on existing smartphones can be used.}
% \vspace{-0.2in}
\label{fig:distances_phones}
\end{figure}

\vskip 0.05in\noindent{\bf Effect of distance.} We evaluate the effect that distance between the sample and the LiDAR and camera has on viscosity coefficient. We vary the distance 5~cm increments from 5 to 20~cm. As before, we run experiments with 0\% fat milk, whole milk and cream each with a volume of 20~\textmu l. At these distances we use a zoom factor of 8x. As shown in Fig.~\ref{fig:distances_phones}, we find that a close distance of 5~cm is needed to obtain the expected correct order of viscosity. As expected, the nearest distance is where the speckle is most clear, and variations in movement can be more easily picked up by the receiving camera. At further distances it becomes challenging to correctly order the viscosity from the received speckle pattern. As the sample is moved further away, the received videos suffers from two forms of attenuation. Firstly, the amount of LiDAR power impinging on the sample is attenuated due to increased distance. Secondly, the speckle reflection is attenuated on the path back to the receiving camera.

\vskip 0.05in\noindent{\bf Using different phone cameras.} Finally, we evaluate how different smartphone cameras  are capable of picking up laser speckle patterns generated by the iPhone LiDAR. Note that we follow well-known procedures to remove the near infrared filter on various smartphone models~\cite{eigen}. To demonstrate that we do not need high end cameras, we perform this benchmark with cameras on an iPhone 5s, Samsung Galaxy S4 and Samsung Galaxy Note 10+, which spans phones from 2013 to 2019. We repeat the above experiments where the iPhone LiDAR and the smartphone cameras are co-located at the same distance of 5~cm from the fluid sample. We use the highest zoom factor available on these smartphones. Fig.~\ref{fig:distances_phones} shows that across these phone cameras, we were able to  correctly capture the viscosity ordering for the dairy products. This suggests that even the cameras on older phones have sufficiently high resolution to be able to discern speckle images and distinguish at least between coarse grades of viscosity. Measurement differences between phone camera models can be accounted for with a calibration step to map the coefficient values to the  ground truth.

\section{Related work}
\label{sec:related}

Prior work can be broadly divided into four different domains (Table~\ref{table:related}).

\vskip 0.05in{\bf Liquid sensing.} There has been recent interest in the mobile computing community on sensing the properties of liquids using wireless signals. LiquID~\cite{liquid} uses ultra-wide band (UWB) radios to measure the permittivity of the liquid by placing it between a pair of UWB transmitter and receiver. TagScan~\cite{tagscan} identifies different material by using RFID tags and by analyzing the phase and received signal strength changes as the radio signal penetrates through the target material. RFIQ~\cite{rfiq} uses radio coupling of RFID tags with the material to classify different materials. Similarly, RadarCat~\cite{radarcat} uses Soli radar to identify various materials. All these systems leverage radio signals and are complementary to the laser speckle based design explored in this paper. Furthermore, these systems are designed for liquids and material that have orders of magnitude larger volume than the tens of microliters that our system  operates with.

Prior work has also used the photoacoustic effect to identify contaminants in liquids (e.g., milk)~\cite{nutrilyzer}. These systems use custom hardware to listen to the sounds generated by intensity modulated light while passing through liquids. This requires custom photoacoustic hardware and is also designed to operate with tens of milliliters of liquid which is three orders of magnitude greater volume than our target of a drop of liquid. In contrast, we explore the use of LiDAR hardware that are been recently incorporated into smartphones  as a  source of laser to create a laser speckle  system.

Prior work also capture the liquid surface waves using the smartphone camera and compute the surface tension on the liquid~\cite{capcam_0,capcam}. Specifically, systems such as CapCam~\cite{capcam} use the vibration motor on the smartphone to create capillary waves in the liquid that is in a paper cup. These waves are illuminated using the flashlight on the phone and used to compute the surface tension.  These systems require visibly creating capillary waves in the liquid and require substantial amount of liquid to work. Specifically, the system fails to compute the surface tension when the depth of the liquid in a standard paper cup is less than 25 mm~\cite{capcam}. Prior work using pendant drop methods are capable of computing surface tension~\cite{pendantdrop_0,pendantdrop_1}. To do this, a drop of liquid is suspended from a tube where the drop's shape is determined by the surface tension and the weight of the drop. As a result this requires custom hardware to tightly control the drop's size and shape. In contrast, the laser speckle based approach works across different liquid volumes between 10-50~\textmu l (see~\xref{sec:benchmark}).

\begin{table}[h!]

\caption{Comparison of related sensing technologies. Related systems have been used for liquid identification, detection, identifying parameters related to nutrients, measurement of physical quantities including permittivity and surface tension, while our work is focused on liquid classification.}

\footnotesize
\begin{adjustwidth}{-0.2cm}{}
\centering
\begin{tabular}{ |M{7em}|M{7em}|M{10em}|c|M{6em}|c|c| } \hline
{\bf System}&{\bf Technology}&\thead{Quantity\\measured}&\thead{Smartphone\\sensors only}&\thead{Liquid\\ classification}&\thead{Min. liquid\\volume}&{\bf Contactless?} \\ \hline

LiquID~\cite{liquid} & Ultra-wideband radios & Permittivity & \rcross. & \gcheck. & > 1 cup & \gcheck. \\ \hline
TagScan~\cite{tagscan} & RFID & Phase and received signal strength of radio signal & \rcross. & \gcheck. & $\sim$300~ml & \gcheck. \\ \hline
RFIQ~\cite{rfiq} & RFID & Amplitude and phase of reflected radio signal & \rcross. & \gcheck. & $\sim$1 cup & \gcheck. \\ \hline
RadarCat~\cite{radarcat} & Radar (FMCW, 57-64 GHz) & Reflected radio signals  & \rcross. & Detection only & $\sim$1 cup & \rcross. \\ \hline
Nutrilyzer~\cite{nutrilyzer} & Photoacoustic effect & Photoacoustic spectra & \rcross. & \gcheck. & $\sim$4.5~ml & \gcheck. \\ \hline
CapCam~\cite{capcam}  & Smartphone vibration motor and camera & Surface tension & \gcheck. & \gcheck. & $\sim$1 cup & \gcheck. \\ \hline
~\cite{capcam_0} & External vibration source and smartphone camera & Surface tension coefficient & \rcross. & \gcheck. & 4.8~l & \gcheck. \\ \hline
~\cite{pendantdrop_0,pendantdrop_1} & Pendant drop & Surface tension & \rcross. & \gcheck. & 1 drop & \gcheck. \\ \hline

~\cite{zuniga2021ripe} & Green light (LED) sensing & Reflected green light  & \rcross. & \rcross. & N/A & \rcross. \\ \hline
Deep thermal imaging~\cite{cho2018deep} & Thermal imaging & Thermal images & \rcross. & \rcross. & N/A  & \gcheck. \\ \hline
MIDAS~\cite{emenike2021characterizing} & Thermal imaging & Thermal dissipation time & \gcheck. & \rcross. & N/A & \gcheck. \\ \hline

~\cite{kira2016detection} & Open path Fourier-transform infrared spectroscopy (OP-FTIR) & OP-FTIR spectral response & \rcross. & \gcheck. & Water droplets & \gcheck.\\ \hline
~\cite{bello2020nir} & Near-infrared spectroscopy & Near infrared spectral response & \rcross/ & \gcheck. & 18~\textmu L & \gcheck. \\ \hline

% ~\cite{harvard1,blood2} & Laser speckle phenomena & Prothromin (clotting) time international normalized ratio & \rcross. & N/A & 40 -- 100~\textmu L & \gcheck. & Closed form equation\\ \hline
% ~\cite{dairylaser} & Laser speckle phenomena & Viscosity index & \rcross. & \gcheck. & Unspecified & \gcheck. & Closed form equation\\ \hline
{\bf Laser speckle using smartphone LiDAR (this work)} & {\bf Laser speckle phenomena} & {\bf Viscosity index} & \gcheck. & \gcheck. & {\bf 10~\textmu L} & \gcheck. \\ \hline

\end{tabular}
\label{table:related}
\end{adjustwidth}
\end{table}

\vskip 0.05in{\bf Object sensing.} Beyond liquid sensing, there has also been recent interest in leveraging sensors on smartphones and smartwatches for sensing object and material properties. Optical sensing solutions that leverage the reflections of green light have been proposed for tracking the quality of food and produce over time~\cite{zuniga2021ripe}. The system relies on two green LED lights and a photoreceptor that are present on some smartwatches to measure the reflection from an object. However, the system is not designed to work with liquids, and has to be in physical contact with the object. Thermal imaging has been used for classifying between different surfaces and objects. Deep Thermal Imaging~\cite{cho2018deep} uses thermal images obtained from a smartphone-attachable thermal camera to classify between different surfaces and materials. MIDAS~\cite{emenike2021characterizing} uses a smartphone with an onboard thermal camera to measure the change across thermal images after a subject has touched an object, and trains a classifier to distinguish between different objects. However, none of these solutions are designed for liquid sensing and are instead designed for classification of larger solid objects or surfaces. Our work is focused on sensing of liquids and uses the laser speckle phenomena which is capable of making distinctions between small quantities of liquid samples.
 
\vskip 0.05in{\bf Spectral analysis techniques.} Laboratory techniques such as Fourier transform infrared spectroscopy (FTIR) including open path FTIR (OP-FTIR) and near-infrared spectroscopy (NIRS) are used to measure the spectral response of a sample in terms of its transmittance or reflectance across different wavelengths~\cite{owen1999near,movasaghi2008fourier}. These techniques have numerous applications including measurement of cerebral blood flow and volume~\cite{owen1999near}, identification and quality analysis of pharmaceutical substances~\cite{luypaert2007near}, and polymer identification for microplastics analysis~\cite{hanvey2017review}. FTIR techniques have been used to measure and quantify the amount of water droplets and other solutes sprayed into the air~\cite{kira2016detection}. NIRS can be used to distinguish between liquids like water, alcohol, benzene and other chemicals at volumes as low as 2.5\textmu l in a capillary tube~\cite{murayama2003near,bello2020nir}. NIRS has also been used to extract a liquid's fat, protein, and lactose content from milk~\cite{vsavsic2001short}. NIRS devices exist in portable form factors and have been developed to either attach~\cite{huang2021applications} or wirelessly connect to~\cite{trinamix} smartphones. The company FrinGOe~\cite{fringoe} has a patent~\cite{gaiduk2019image} for a compact FTIR spectroscope, and a smartphone case that can perform spectroscopy in the visible light range of 400 -- 700~nm wavelength range, however these products are not commercially available. In contrast to using expensive desktop laboratory equipment or developing costly attachments, our work seeks to leverage sensors available on existing smartphones to perform liquid sensing.

\vskip 0.05in{\bf Laser speckle techniques.} The theory behind the speckle phenomena  was developed in 1960s as laser hardware became more common in laboratories~\cite{laser1,survey}. In 1970s, researchers started exploring the time-varying nature of the speckle in the presence of motion such as the movement of red blood cells~\cite{laser2}. Single-exposure speckle photography was developed to study the blood flow in retinas in the 1980s~\cite{laser3}. As digital photography became commercially possible in the 1990s, speckle analysis became more real-time and practical~\cite{survey}. In the last decade, a number of  applications of laser speckle have been demonstrated in the biomedical research community~\cite{clinicalsurvey}. This includes the use of laser speckle to monitor the dynamic vascular reactivity in systemic sclerosis patients~\cite{lasercapp1}, blood coagulation analysis~\cite{harvard1,blood2}, burn wound assessment~\cite{lasercapp2}, monitoring birthmarks~\cite{lasercapp3}, dental decay~\cite{lasercapp5} and a variety of opthalmic conditions such as glaucoma, retinopathy, and macular degeneration~\cite{lasercapp4}.

Dynamic laser speckle has also been used for non-medical applications such as imaging the dynamic properties of leaf topography on the scale of the wavelength of laser light~\cite{laserplant}, to compute the viscosity of dairy products~\cite{dairylaser} and for the detection of flat, curved, smooth, and rough microplastics in water~\cite{laserplastics}. These prior work, however use expensive and custom hardware with high resolution cameras and high power lasers that is not commonly available. Recent efforts have tried to create smartphone attachments of laser hardware that use the camera on the smartphone and a custom visible light laser hardware to capture the laser speckle reflections~\cite{phonelaser1,phonelaser2}. In contrast to using custom laser hardware, our paper introduces the idea of repurposing the LiDAR hardware recently been added to smartphones (e.g., iPhone Pro) as an near-infrared laser source to achieve laser speckle reflectometry. We show various proof-of-concept applications using off-the-shelf smartphones and show that the LiDAR hardware in smartphones can be useful for both biomedical sensing and food rheology applications.

\section{Discussion and Conclusion}
We introduce the idea of repurposing smartphone LiDAR for liquid sensing by creating a laser speckle system. We evaluate our design with two application including sensing fat content and adulterants in milk as well as identifying coagulated blood. Here describe the current limitations of our design as well as future research opportunities.

\vskip 0.03in\noindent{\bf Unmodified near-infrared camera.}  In our evaluations, we remove the near infrared filter from a cheaper smartphone in order to capture speckle patterns. However, the iPhone has an onboard near-infrared camera and it is conceivable that more phone manufacturers will provide software access to it in the future, as well as more granular control on image settings. In particular, being able to have greater control over image quality, shutter speed and frame rate could be used to increase the viscosity resolution of mobile laser speckle sensing systems such as being able to distinguish between fine-grained dilutions of milk.

\vskip 0.03in\noindent{\bf Closed-form viscosity equation across all liquids.} Laser speckle is formed due to constructive and destructive interference patterns from minute particles that are larger than the size of the laser wavelength. Since the Brownian motion of the particles is affected by its viscosity, it can be used to get a proxy measure of viscosity by measuring the speckle pattern. However, the laser speckle is dependent on the particles in the liquid and its viscosity. For example, particles in blood like red blood cells and platelets are different from particles in say sparkling water. So, getting a closed-form equation for viscosity across all types is liquids is challenging and thus the machine learning based approach in this paper is more compelling.

\vskip 0.03in\noindent{\bf LiDAR transmission power.} Typical laser speckle sensing systems use lasers that with an output power in the range of 5--20~mW. While we do not know the exact output power from the iPhone LiDAR scanner, the image intensity of objects illuminated with it is substantially lower than when illuminated with a 5~mW near infrared laser diode. One limitation of the low-power LiDAR transmitter is that when it is pointed at a transparent liquid like water, the laser speckle pattern that dominates is the one scattered from the surface itself, instead of the liquid, which is much weaker and difficult to image. As such, a higher power laser would be needed to sense viscosity of transparent or translucent fluids. We also note that laser sensing systems often use polarizers in order to capture the cross-polarized laser speckle pattern, which is free of interference from the transmitter. Polarizers are cheap, passive components which can improve the signal-to-noise ratio of the received speckle pattern.

\vskip 0.05in\noindent{\bf Caveats with using viscosity for fluid sensing.} There are a couple of caveats of using any kind of viscosity measurement tool to identify fluid composition. First, the temperature affects the viscosity of a liquid. Thus, to use viscosity measuring tools to identify fat content or adulteration requires that the experiments are run under a standardized temperature. Second, viscosity can be used to either identify deviations from known fluids or distinguish between the viscosity of known fluids. Identifying the chemical composition of an unknown adulterant in the fluid is challenging with this technique. Given the widespread consumption of milk, we envision that our smartphone LiDAR approach can be used as a screening tool to identify deviations from expected viscosity values and further testing might be required to identify the chemical composition of unknown adulterants. 

\vskip 0.03in\noindent{\bf Exploring other applications.} As described in~\xref{sec:related}, laser speckle imaging has been demonstrated to be useful for various bio-medical applications including burn wound assessment~\cite{lasercapp2}, monitoring birthmarks~\cite{lasercapp3}, dental decay as well as for the study of blood flood in retinas~\cite{laser3}. Exploring these and other non-biomedical applications like analysing micro-plastics~\cite{laserplastics} would be impactful future research directions for our accessible smartphone-LiDAR based system.

\section*{Acknowledgments}
The authors were funded in part by awards from the National Science Foundation and Moore Inventor Fellow. We would like to thank Joanne K Estergreen at the Hematology and Coagulation Lab and Department of Laboratory Medicine and Pathology at the University of Washington Medical Center.

\bibliographystyle{ACM-Reference-Format}
\bibliography{ms}

\end{document}